\documentclass[12pt]{iopart}   
\usepackage{graphicx}
\usepackage{epsfig}
\usepackage{amssymb}
\begin{document}

\newcommand{\be}{\begin{equation}}
\newcommand{\ee}{\end{equation}}
\newcommand{\beqn}{\begin{eqnarray}} 
\newcommand{\eeqn}{\end{eqnarray}}

\title[Extinction transition in stochastic population dynamics]{Extinction transition in stochastic population dynamics in a random, convective environment}

\author{R\'obert Juh\'asz}
\address{Institute for Solid
State Physics and Optics, Wigner Research Centre for Physics, H-1525 Budapest,
P.O. Box 49, Hungary}
\ead{juhasz.robert@wigner.mta.hu}

\begin{abstract}
Motivated by modeling the dynamics of a population living in a flowing medium where the environmental factors are random in space, 
we have studied an asymmetric variant of the one-dimensional 
contact process, where the quenched random reproduction rates are systematically greater in one direction than in the opposite one. 
The spatial disorder turns out to be a relevant perturbation but, according 
to results of Monte Carlo simulations, the behavior of the model at the extinction transition is different from the (infinite randomness) critical behavior of the disordered, symmetric contact process. 
Depending on the strength $a$ of the asymmetry, the critical population drifts either with a finite velocity or with an asymptotically vanishing velocity as $x(t)\sim t^{\mu(a)}$, where $\mu(a)<1$. 
Dynamical quantities are non-self-averaging at the extinction transition; 
the survival probability, for instance, shows multiscaling, i.e. it is
characterized by a broad spectrum of effective exponents. 
For a sufficiently weak asymmetry, a Griffiths phase appears below the extinction transition, where the survival probability decays as a non-universal power of the time while, above the transition, another extended phase emerges, where
the front of the population advances anomalously with a diffusion exponent continuously varying with the control parameter. 
\end{abstract}

\maketitle

\section{Introduction}
A fundamental question of population dynamics is the long-term fate of a population --- extinction or survival.  Which one of these possibilities occurs, 
is controlled by the reproduction and death rates of individuals, 
and through them, it is influenced by many factors of the environment such as the availability of nutrients, etc. 
A paradigmatic model in this field is the {\it contact process} \cite{harris,liggett}. In this model, individuals placed on the sites of a lattice (at most one per lattice site) either produce an offspring on a neighboring empty site or die stochastically, with certain rates. 
Tuning these rates, the model undergoes a continuous phase transition from a fluctuating survival phase to an extinction phase, also termed as {\it absorbing phase transition} \cite{md,hhl}. 
The phase transition of the spatially homogeneous variant of the model falls into the universality class of {\it directed percolation} \cite{md,hhl,odor}. 
At the phase transition, various observables like the survival probability or the number of individuals follow power laws and the corresponding critical dynamical exponents are known with high precision. 

In reality, environmental factors are not homogeneous in space;
therefore it is an important question, whether an extinction transition is stable against spatial heterogeneity and if not, what characteristics it has. 
Spatial heterogeneity can be taken into account in the contact process by allowing for site-dependent random reproduction and death rates. 
It turned out that the directed percolation universality class is unstable against this type of disorder \cite{noest,janssen}, and the critical dynamical observables scale as a power of the logarithm of the time, which is termed as {\it activated scaling} \cite{moreira}. This has been confirmed, at least for strong enough disorder, by a renormalization group method, which describes the critical behavior by a so called infinite-randomness fixed-point \cite{im} and provides the complete set of critical exponents in one dimension \cite{hiv}. 
Whether the activated scaling is valid for an arbitrarily weak disorder is an open question \cite{hiv,vd,nft,hoyos}. 
Besides the ultra-slow critical dynamics, the behavior of the system is anomalous in the extinction phase, as well. Here, in the {\it Griffiths phase} 
\cite{griffiths}, the
time-dependence of average quantities follows power laws with exponents varying continuously with the control parameter.  
This is due to the presence of spatial domains with favorable environmental factors (i.e. high reproduction rate), which give a large contribution to the average \cite{noest,cgm}. 

Recently, several works have been devoted to the effect of a flowing medium (``wind'') on the dynamics of a population in a heterogeneous environment
\cite{nelson,lebowitz,kessler,neicu}.
In addition to reproduction and death, individuals were allowed to drift in a given direction in the corresponding models and   
heterogeneity was represented by a single domain with favorable environmental factors (``oasis'') in a ``desert'' \cite{kessler,lebowitz,neicu} or by a random environment \cite{nelson}. The focus was on a localization-delocalization transition when the drift velocity is varied \cite{nelson}. 

Note that the effect of convection in a translationally invariant environment is trivial; using a deterministic model such as the Fisher-Kolmogorov equation \cite{saarloos}, the convection term can be canceled by switching to a co-moving frame. The phase transition of the corresponding stochastic model, the asymmetric contact process, which has different reproduction rates in the two directions \cite{acp}, still falls in the directed percolation universality class \cite{tretyakov}.   

The aim of this work is to study population dynamics in a random environment and in the presence of convection at the absorbing transition. 
For this purpose, a deterministic description such as the spatially heterogeneous generalization of the Fisher-Kolmogorov equation analysed in Ref. \cite{nelson} is inappropriate since, close to the extinction transition, fluctuations play an important role. 
Therefore we will consider a stochastic model, the one-dimensional, asymmetric contact process with random transition rates, where the reproduction occurs
with systematically larger rates in a given direction than in the opposite one. 
A recent, strong disorder renormalization group treatment of this model concluded that the absorbing phase transition of the model is out of the validity of the method and, consequently, the scaling is not of activated type \cite{juhasz}.  
In this work, we will present results of Monte Carlo simulations in order to gain insight into the critical behavior.  

The rest of the paper is organized as follows. In section \ref{model}, the definition and some basic properties of the model are given.
The density profile induced by an active boundary is calculated within a mean-field approximation in section \ref{meanfield}. 
Results of Monte Carlo simulations are presented in section \ref{tascp} for the totally asymmetric model and in section \ref{pascp} for the partially asymmetric model. Section \ref{discussion} is devoted to the discussion of the results and some calculations are presented in the appendix.

\section{The model and its basic properties}
\label{model}

\subsection{Definition of the model} 

Let us have a one-dimensional lattice, the sites of which are either 
occupied by an individual or empty, and consider a continuous-time Markov process with the following independent transitions. 
Individuals produce offsprings on the adjacent site on their left and right with site-dependent rates $\kappa_i$ and $\lambda_i$, respectively, or die with site dependent rates $\mu_i$. This model with general random transition 
rates has been studied in Ref. \cite{juhasz}. 
We will consider here the special case, where the ratios $\kappa_i/\lambda_i$ of reproduction rates are constant, and define the asymmetry parameter $a$ as $\kappa_i/\lambda_i=a/(1-a)$ .
The case $a=1/2$ corresponds to the (disordered) contact process while, for 
$a\neq 1/2$, the asymmetry induces a drift of the population in space in either direction.  
Note that the drift in this model is not the result of the motion of individuals (which are immobile here) like in the contact process with Kawasaki exchange kinetics \cite{lebowitz} but it is a consequence of the asymmetry of reproduction process. Nevertheless, on large scales, the differences between the two cases are expected to be irrelevant; one can show that, in the continuum mean-field limit, both models are described by a disordered variant of the Fisher-Kolmogorov equation.  

In the Monte Carlo simulations, we have implemented a discrete time version of the above process as follows. 
An individual is chosen randomly, which either dies with the site-dependent 
probability $1/(1+r_i)$ on site $i$ or, with the probabilities $ar_i/(1+r_i)$ and $(1-a)r_i/(1+r_i)$, tries to produce a new individual on the adjacent site on its left and right, respectively. 
One Monte Carlo step (of unit time) consists of $N(t)$ such updates, where $N(t)$ is the number of individuals at the beginning of the Monte Carlo step. 
The parameters $r_i$ are i.i.d. quenched random variables drawn from a discrete distribution with the probability density 
\be 
f(r)=c\delta(r-\lambda) + (1-c)\delta(r-w\lambda), 
\label{fr}
\ee  
where $0\le c\le 1$ and $0<w\le 1$ are parameters and $\delta(x)$ is the Dirac delta distribution.
In words, on a fraction $c$ ($1-c$) of sites, the reproduction rate is high (low) and the death rate is low (high). 
The parameter $c$ has been set to $c=1/2$ throughout the simulations, and the remaining three parameters of the model are the control parameter $\lambda$, the asymmetry $a$ and the strength of disorder $w$ (the case $w=1$ corresponds to the homogeneous model). In the following we assume that $0\le a<1/2$, meaning that the population drifts from left to right.      

\subsection{Quantities of interest}

We have started the simulations from an initial state where
a single individual is placed on an otherwise empty (infinitely large) lattice.
We were then interested in the following quantities, all defined in a given random environment (i.e. set of parameters $\{r_i\}$) and 
for a given starting position.  
The survival probability $P(t)$, which is the probability that there is at least one individual on the lattice at time $t$, the expected number $N(t)$ of individuals and, respectively, the expected position $x_l(t)$ and  $x_r(t)$ of the leftmost and rightmost individuals at time $t$ (conditioned on survival up to time $t$). 

Beside the survival probability, in the presence of an asymmetry ($a<1/2$) it is also a relevant question how far the population drifts from the origin.  
To characterize this, we define the hitting probability $Q_{0,n}$ as the probability that the population, initiated on site $0$, reaches site $n$ (any time).   

In addition to this, one can imagine a situation that the environment contains a region with highly favorable conditions where the population lives for very long time and which serves as a steady source of individuals for the surrounding region. 
This can be modeled by an {\it active boundary} at site $0$, i.e.
by putting an individual there, whose lifetime is infinite ($\mu_0=0$). 
Then, the process settles in a non-trivial steady state and one may ask what is the probability to find an individual at site $n$ in this state. The set of these probabilities $\rho_n^a$ is called the (stationary) density profile.  

Obviously, all the above quantities will depend on the particular realization of the random environment, and their averages over random environments will be denoted by an overbar.  

An alternative initial condition often used for probing the dynamics is a state where all sites are occupied. One is then interested in the dependence of the density $\rho_n(t)$ on time. 
For the symmetric contact process, this problem is exactly related to the single seed initial condition, due to a property of the model called self-duality \cite{liggett}. Namely, the density $\rho_n(t)$ for the fully occupied initial condition is equal to the survival probability $P_n(t)$ in the process started from a single individual on site $n$. Using the quantum Hamilton formalism, this relation can be shown to hold even in the disordered model, if the processes for the two different initial conditions are considered in the same random environment \cite{hv}.   
We will show in the appendix that, for the asymmetric contact process, a generalized relationship 
\be 
\rho_n(t;\Omega)=P_n(t;\tilde\Omega). 
\label{rhoP}
\ee
exists, which relates the observable  $\rho_n(t)$ 
in a random environment $\Omega\equiv\{\lambda_i,\kappa_{i+1},\mu_i\}$ to $P_n(t)$ in the {\it dual} environment $\tilde\Omega\equiv\{\kappa_{i+1},\lambda_i,\mu_i\}$ obtained from the original one by interchanging $\lambda_i$ and $\kappa_{i+1}$ on each link.
If the probability measure for random environments is invariant under the duality transformation and a subsequent reflection $n\to -n$, which is the case for independent rates, the averages over the random environments will be equal in an infinite system: 
\be 
\overline{\rho}(t)=\overline{P}(t). 
\label{rhoPav}
\ee

The relationship in Eq. (\ref{rhoP}) holds in the case of an active boundary ($\mu_0=0$), as well. This enables us to obtain another relationship between the stationary profile and the hitting probability, as follows. 
The local densities $\rho_n(t;\Omega)$ when started from the fully occupied state tend to the stationary profile $\rho_n(\Omega)$ as $t\to\infty$. 
Using Eq. (\ref{rhoP}), $\rho_n(\Omega)=\lim_{t\to\infty}P_n(t;\tilde\Omega)$. 
But the probability that the population starting from site $n$ survives in limit $t\to\infty$ must be be equal to the probability that it reaches the origin (where the death rate is zero) at some time. 
So we have 
\be
\rho_n^a(\Omega)=Q_{n,0}(\tilde\Omega), 
\ee
and, if the probability measure of random environments has the property described above Eq. (\ref{rhoPav}), then 
\be 
\overline{\rho_n^a}=\overline{Q_{0,n}}.
\ee
In the following, we will the omit the index referring to the origin and denote the hitting probability simply by $Q_n$. 

\subsection{Limiting cases}

In this section, we shall survey the critical properties of two limiting cases of our model, the clean contact process ($w=1$) and the disordered, symmetric contact process ($a=1/2$), most of which are known from previous works. 

In the homogeneous contact process, approaching the critical point $\lambda_c$ from above, the order parameter $P(\infty)\equiv\lim_{t\to\infty}P(t)$ vanishes as 
$P(\infty)\sim (\lambda-\lambda_c)^{\beta}$ and 
the temporal and spatial correlation lengths diverge as 
$\tau\sim (\lambda-\lambda_c)^{-\nu_{\parallel}}$ and 
$\xi\sim (\lambda-\lambda_c)^{-\nu_{\perp}}$, respectively. 
The survival probability, the mean number of particles and 
the spatial extension 
$l(t)\equiv x_r(t)-x_l(t)$ of the population behave at the critical point 
$\lambda=\lambda_c$ as 
\beqn
P(t)\sim t^{-\delta}, \label{Pt} \\
N(t)\sim t^{\eta},  \\
l(t) \sim t^{1/z}, \label{lt}
\eeqn 
irrespective of $a$. 
The values of critical exponents can be found in Table \ref{table}. 
%%%%%%%%%%%%%%%%%%%%%%%%%%%%%%%%%%%%%%%%%%%%%%%%%%%%%%%%%%%%%%%%%%%
\begin{table}[h]
\begin{center}
\begin{tabular}{|l||l|}
%\hline   & CP  \\
%\hline
\hline  $\lambda_c$       &  3.297848(22)\\
\hline  $\delta$          &  0.159464(6) \\
\hline  $\eta$            &  0.313686(8) \\
\hline  $z$               &  1.580745(10)\\
\hline  $\nu_{\parallel}$   &  1.733847(6) \\
\hline  $\nu_{\perp}$      &  1.096854(4)  \\
\hline  $\beta$           &  0.276486(8)  \\
\hline
\end{tabular}
\end{center}
\caption{\label{table} 
The location $\lambda_c$ of the critical point \cite{hhl} and the critical exponents \cite{jensen} of the homogeneous, symmetric contact process.}
\end{table}
%%%%%%%%%%%%%%%%%%%%%%%%%%%%%%%%%%%%%%%%%%%%%%%%%%%%%%%%%%%%%%%%%%%%%%
The position-dependent probabilities $Q_n$ and $\rho_n^a$ behave differently for $a=1/2$ and $a<1/2$ at the critical point. 
Starting the process from a single individual in the origin and 
assuming that it survives up to time $t$, the rightmost individual will typically be on site $n(t)\sim t^{1/z}$ if $a=1/2$, and on site $n(t)\sim t$ if $a<1/2$, where we have used Eq. (\ref{lt}) for $a=1/2$ and that the population drifts with a constant velocity for $a<1/2$. 
From the relation $P(t)\sim Q_{n(t)}$ and Eq. (\ref{Pt}), we obtain 
\beqn
Q_n\sim n^{-z\delta}\sim n^{-\beta/\nu_{\perp}} \quad  {\rm if} \quad a=1/2, 
\label{Qna}\\
Q_n\sim n^{-\delta}\sim n^{-\beta/\nu_{\parallel}} \quad {\rm if} \quad a<1/2.
\label{Qnb}  
\eeqn
Here, we have applied the scaling relations $\delta=\beta/\nu_{\parallel}$ and 
$z=\nu_{\parallel}/\nu_{\perp}$ of directed percolation, see e.g. Ref. \cite{md}. 
The same results as given in Eqs. (\ref{Qna}-\ref{Qnb}) have been obtained 
for the stationary profile $\rho_n^a=Q_n$ by scaling arguments in Ref. \cite{costa}. 

The critical behavior of the disordered, symmetric contact process ($a=1/2$, $w<1$) is much different from that of the homogeneous model. 
Here, a strong disorder renormalization group method predicts that the average dynamical quantities, which are dominated by $O(1)$ contributions of rare, 
atypical environments, behave at criticality as 
\beqn
\overline{P}(t)\sim (\ln t)^{-\tilde{\delta}},  \label{actP} \\
\overline{N}(t)\sim (\ln t)^{\tilde{\eta}},     \label{actN} \\
\overline{l}(t) \sim (\ln t)^{1/\psi},        \label{actl}
\eeqn 
where the exponents $\tilde{\delta}=(3-\sqrt{5})/2$, $\tilde{\eta}=\sqrt{5}-1$ and $\psi=1/2$ calculated by the method are conjectured to be exact \cite{hiv}. This behavior has been confirmed by large scale simulations even for relatively weak disorder \cite{vd}.  
In such an infinitely disordered critical point, the dynamical exponents $\delta$, $\eta$ and $z$ defined through algebraic relations as given in Eqs. (\ref{Pt}-\ref{lt}) are formally $\delta=\eta=1/z=0$.
By scaling arguments, we obtain that the average probability $\overline{Q_n}=\overline{\rho^a_n}$ decays at the critical point as 
\be 
\overline{Q_n}=\overline{\rho^a_n}\sim n^{-\tilde{\beta}/\tilde\nu_{\perp}}, 
\ee
where the exponent takes the value $\tilde{\beta}/\tilde\nu_{\perp}=(3-\sqrt{5})/4$ \cite{hiv}.

\subsection{Absorbing phase transition in the disorder, asymmetric model} 

As a first step in studying the effects of a random environment, one can investigate whether the phase transition of the homogeneous model is stable against a {\it weak} disorder. This can be done by a suitable formulation of the heuristic argument by Harris \cite{h}, as follows. 
Due to disorder, the variation of the average control parameter in a domain of size $L$ is $\Delta(L)\sim L^{-1/2}$, according to the central limit theorem. 
The corresponding correlation length at a distance $\Delta(L)$ from the critical point is $\xi\sim[\Delta(L)]^{-\nu}\sim L^{\nu/2}$.
Clearly, the effect of disorder is irrelevant if $\xi\gg L$ in the limit $L\to\infty$, which occurs if $\nu>2$. 
In the opposite case, the correlations are determined by the disorder and the transition is expected to be either smeared or different from that of the homogeneous system. 
Substituting $\nu=\nu_{\perp}$ in the above relations which is valid for the symmetric contact process, yields the well-known result that 
disorder is relevant ($\nu_{\perp}<2$) \cite{noest}. 
In case of the asymmetric contact process, it was argued in Ref. \cite{costa} that even the spatial correlations diverge with the temporal correlation length exponent $\nu_{\parallel}$ of the symmetric model. 
Since $\nu_{\parallel}<2$ (see Table \ref{table}), also the asymmetric model is expected to be unstable against weak disorder. 
We will see in section \ref{tascp} that Monte Carlo simulations confirm this prediction.   

In the remaining part of the section, we shall show that, for a discrete, binary distribution of parameters $r_i$ given in Eq. (\ref{fr}), two qualitatively different kinds of phase transition can be distinguished. 

Our starting point is that a random environment consists alternately of segments with favorable sites (termed as sites of type A) and segments of unfavorable sites (sites of type B) of variable lengths, which follow a geometric distribution with the parameter $c$. 
Let us consider now a homogeneous, infinite environment that consists exclusively of sites of type A. 
There are two distinct phase transitions in this model \cite{acp}. 
An absorbing phase transition at $\lambda=\lambda_1^0(a)$, below (above) which the order parameter $P(\infty)$ is zero (positive), and another one at a higher value $\lambda_2^0(a)$, below (above) which the probability of finding an individual in the origin in the limit $t\to\infty$ is zero (positive).
In other words, in the domain  $\lambda_1^0(a)<\lambda<\lambda_2^0(a)$, which we call the {\it weak-survival phase}, both the mean positions of the leftmost and the rightmost individuals of the population (provided that it survives up to time $t$) shift rightwards with a constant velocity, i.e. $\overline{x_l}(t)\sim \overline{x_r}(t)\sim t$. 
Whereas, in the domain $\lambda>\lambda_2^0(a)$, termed as the {\it strong-survival phase}, the right end of the population propagates rightwards as before but the left end moves leftwards (with a constant velocity).  
(If  $\lambda<\lambda_1^0(a)$, the population gets extinct in the limit 
$t\to\infty$; this domain is the extinction phase.)
In the symmetric model the two phase transitions coincide, i.e. $\lambda_1^0(1/2)=\lambda_2^0(1/2)$. 
The location of the lower transition point varies weakly with $a$; starting from the value given in Table \ref{table}, which is valid for $a=1/2$, 
it goes continuously to 
$\lambda_1^0(0)=3.306(2)$ of the totally asymmetric model ($a=0$) 
\cite{tretyakov}. As opposed to this, $\lambda_2^0(a)$ increases rapidly with $a$, and diverges as $a\to 0$. 
Obviously, in the totally asymmetric limit, $\overline{x_l}(t)$ is non-decreasing and, therefore, the strong-survival phase is missing.     

Considering a homogeneous, infinite system consisting of sites of type B, evidently, the two transition points are located at $\lambda_1^0(a)/w$  and $\lambda_2^0(a)/w$. 

In case of a finite homogeneous segment of size $L$, the mean lifetime of the population initiated from a single individual on its leftmost site is different in the above three phases.    
In the extinction phase it is finite and, for large $L$, asymptotically independent of $L$. In the weak-survival phase, the population drifts rightwards with a constant velocity and dies out at the right end, so the lifetime is $O(L)$. 
Finally, in the strong-survival phase the population spreads out in the whole segment and reaches the empty state only through large fluctuations of the 
density, which are exponentially improbable in $L$, thus the lifetime is $O(e^{\rm{const}\cdot L})$.

Let us now return to the disordered system. 
Similar to the homogeneous system, one can define here two distinct phase transitions. In this paper, we shall focus on the lower (absorbing) one, which is defined by the vanishing of the average order parameter $\overline{P}(\infty)\equiv \lim_{t\to\infty}\overline{P}(t)$. For the second one, we refer the reader to Ref. \cite{juhasz}. 

If  $\lambda>\lambda_1^0(a)/w$, both segments of type A and B are locally 
in the (either weak or strong) survival phase, therefore the system must be in its survival phase, i.e. $\overline{P}(\infty)>0$. 
Contrary to this, if $\lambda<\lambda_1^0(a)$, both types of segments are in their extinction phase, thus the disordered system is in that phase, as well, i.e.  $\overline{P}(\infty)=0$.
We conclude therefore, that the critical point of the disordered system must be in the range 
\be 
\lambda_1^0(a)<\lambda_c(w,a)<\lambda_1^0(a)/w.
\ee
In this domain, segments of type B are locally in the extinction phase, while
segments of type A are either in the weak or the strong survival phase, depending on $a$,$w$, and $\lambda$. 
If  
\be 
\lambda_1^0(a)<\lambda_c(w,a)<\lambda_2^0(a),
\ee
segments of type A are locally in their weak-survival phase at the phase transition (and below). In this case, we will speak of a phase transition of type I. 
Using the behavior of mean lifetime presented above, we can see that, in the critical point, the left end of the population moves with a constant velocity (in surviving populations), i.e. $\overline{x_l}(t)\sim t$. Since $x_l(t)\le x_r(t)$ and $\overline{x_r}(t)$ is at most linear in $t$, we have $\overline{x_r}(t)\sim t$. Below a phase transition of type I, $\lambda<\lambda_c(w,a)$, the survival probability decays exponentially in time.
 
We have a qualitatively different situation if
\be 
\lambda_2^0(a)<\lambda_c(w,a).
\ee
Here, segments of type A are in their strong-survival phase at the phase transition point of the disordered system. 
In this case, we will speak of a phase transition of type II. 
Below this transition point, in the range $\lambda_2^0(a)<\lambda<\lambda_c(w,a)$, the lifetime of type A segments is exponentially large and, using that their probability of occurrence is exponentially small in their size, we obtain that the survival probability decays here algebraically 
\be 
P(t)\sim t^{-\delta(a,w,\lambda)},
\ee 
with a non-universal decay exponent. This domain is a representant of the well-known Griffiths phase \cite{griffiths,noest}. 
Since, in the critical point, the distribution of the lifetime of (isolated) segments of type A decays still algebraically, the left end (as well as the right end) of the population does not necessarily drift with a constant velocity but it may advance anomalously slowly, as $\overline{x_l}(t)\sim t^\mu$ with $\mu<1$. 
It is the task of the subsequent Monte Carlo analysis to clarify whether such an anomalous drift is indeed realized.  
Below the Griffiths phase,  $\lambda<\lambda_2^0(a)$, $P(t)$ decreases exponentially in time. 

The boundary between the two types of phase transitions in the parameter space 
is given formally by 
\be
\lambda_2^0(a)=\lambda_c(w,a), 
\label{tri}
\ee 
which provides the separatrix $a^*(w)$ in the $w$,$a$ plane. 
For a strong (weak) enough anisotropy such that $a<a^*(w)$  ($a>a^*(w)$), 
the transition is of type I (type II).   
Unfortunately, both sides of Eq. (\ref{tri}) are unknown functions of 
$a$ and $w$. 

\section{Mean-field theory} 
\label{meanfield}

Before presenting numerical results, we will investigate the model 
by a simple mean-field theory, which replaces the joint-probabilities  $P[m_n(t),m_{n+1}(t)]$ of occupation numbers (which are one for occupied and zero for empty sites) by 
$\rho_n(t)\rho_{n+1}(t)$. Thereby the fluctuations of $m_n(t)$ are neglected and the state of the system is described by the set of local densities $\rho_n(t)$. 
Starting from the master equation of the process this leads to the dynamical equations
\be 
\partial_t\rho_n(t)=  -\mu_n\rho_n(t) + [\lambda_{n-1}\rho_{n-1}(t) +
\kappa_{n+1}\rho_{n+1}(t)][1-\rho_n(t)]. 
\label{mfdyn}
\ee
Though this approximation is unable to treat the extinction transition of the model started from a single individual, it gives the behavior of the stationary density profile in the case of an active boundary qualitatively correctly and illustrates the non-self-averaging nature of the disordered model.   
In the followings, the superscript $a$ of the density will be dropped. 

In the mean-field description of the disordered model, the main features of the profile are expected to be independent of the asymmetry. The particular case of total asymmetry ($a=0$) is analytically tractable, so we will restrict ourselves to this case. 
In the stationary state, Eqs. (\ref{mfdyn}) reduce to the recursion equations 
\be
0= -\mu_n\rho_n + \lambda_{n-1}\rho_{n-1}(1-\rho_n), 
\quad n=1,2,\dots
\ee
with the initial condition $\rho_0=1$. 
In terms of the inverse densities $\rho_n^{-1}$ they can be written in a linear form
\be 
\rho_n^{-1}=R_{n-1}\rho_{n-1}^{-1} + 1,
\ee
where $R_{n-1}\equiv \mu_n/\lambda_{n-1}$.  
This recursion is easy to solve and gives 
\be 
\rho_n = [R_{n-1} + R_{n-1}R_{n-2} + \dots +  R_{n-1}R_{n-2}\cdots R_0]^{-1}.
\ee 
The sum in the brackets is known as a {\it Kesten random variable} and appears frequently in one-dimensional disordered systems, see e.g. Ref. \cite{bouchaud}. 
Its asymptotic dependence on $n$ can be easily derived as follows. 
Let us introduce the random variables $\Delta_n\equiv\ln R_n$ and 
$X_t\equiv\sum_{i=1}^t\Delta_{n-i}$, $t=1,2,\dots,n$. 
Since $\Delta_i$ are independent,  $\{X_t\}_{t=1}^n$ is a discrete-time random walk. 
In terms of these variables, $\rho_n$ can be written as 
\be 
\rho_n=\left[\sum_{i=1}^ne^{X_i}\right]^{-1}.
\ee
Two regimes with different asymptotic behavior of $\rho_n$ can be distinguished.If $\overline{\Delta}<0$, the random walk is biased to the negative direction, i.e. $X_n/n\to const<0$ as $n\to\infty$. 
In this case, the Kesten variable converges stochastically and the average density tends to a positive constant, $\overline{\rho_n}\to\rho(\Delta)>0$ as $n\to\infty$.
If, however, $\overline{\Delta}\ge 0$, the Kesten variable diverges and, for large $n$, it is dominated by the maximal term(s) $e^{X_{\rm max}}$, where 
$X_{\rm max}\equiv\max_{i=1,\dots,n}\{X_i\}$.  
If  $\overline{\Delta}>0$, the walk is biased to the positive direction, $X_{\rm max}\sim O(n)$ and the profile decays exponentially, 
$\overline{\rho_n}\sim e^{-n/\xi(\Delta)}$ for large $n$. This means that the population is localized near the active boundary practically within a distance $O[\xi(\Delta)]$. 

At the transition point $\overline{\Delta}\equiv\overline{\ln(\mu/\lambda)}=0$, the typical maximal displacement of the walk is  $X_{\rm max}\sim O(\sqrt{n})$, and the {\it typical} density, i.e. that in almost all realizations of the environment, decreases for large $n$ as 
\be 
\rho_n^{\rm typ}\sim e^{-const\sqrt{n}}.
\ee
The average profile $\overline{\rho_n}$, however, decreases with $n$ much differently, due to a vanishing fraction of atypical realizations of the environment. 
In environments where $X_{\rm max}<0$, namely, the density $\rho_n$ is $O(1)$. 
This condition expresses that the walk never crosses its origin up 
to $n$ steps. 
The probability of this event, termed as persistence probability, is well-known to decay for unbiased random walks as  $P_{\rm pers}(n)\sim n^{-1/2}$ for large $n$. We obtain therefore, that the average density profile decreases algebraically 
as 
\be 
\overline{\rho_n}\sim n^{-1/2}, 
\ee
which is a much slower decay than that in typical environments. 
Though we will see by numerical simulations that the mean-field behavior predicted here is quantitatively incorrect, the feature of the model that typical and average quantities scale differently is captured by this simple theory. 
It is worth mentioning that the average profile in the disordered model decays with a smaller exponent than that of the homogeneous one, where $\rho_n\sim n^{-1}$ \cite{costa}.

\section{Numerical results for the totally asymmetric model}  
\label{tascp}

As we have seen, the model with a sufficiently strong asymmetry, such that 
$a<a^*(w)$ has a phase transition of type I. 
We have performed numerical simulations for a representant of this class, the totally asymmetric model ($a=0$). 

As already mentioned, the different observables of the model depend on the particular realization of the random environment. In order to see the variations caused by the disorder we have measured the mean values in a fixed environment and repeating the measurement in many independent random samples we have constructed the distribution and calculated the typical value, which we define 
as the exponential of the average of the logarithm of the observable. So, for example, the typical survival probability is
\be
P_{\rm typ}(t)\equiv \exp(\overline{\ln P(t;\Omega)}).
\ee 
When calculating the typical observables, the mean value in each environment has been estimated by an average over $10^3$ different stochastic histories, and the subsequent disorder average of the logarithms has been performed over $10^3$ independent realizations of the environment.   
Since sample-to-sample variations of the observables are significant, in the calculations of average quantities, the number of the different runs in a given sample were reduced to one in favor of an increased number of samples, typically $10^5$. 
When measuring of the steady-state profile, the local densities in a given sample were averaged in an interval of $10^5$ Monte Carlo (MC) time steps after a relaxation time of $10^5$ MC steps. 

Simulations have been performed for different values of the strength of the disorder, $w=0.8,0.6,0.4,0.2$. 
The properties of the model for different $w$ were found to be 
qualitatively similar, and detailed results will be presented for $w=0.4$. 

\begin{figure}[h]
\includegraphics[width=0.5\linewidth]{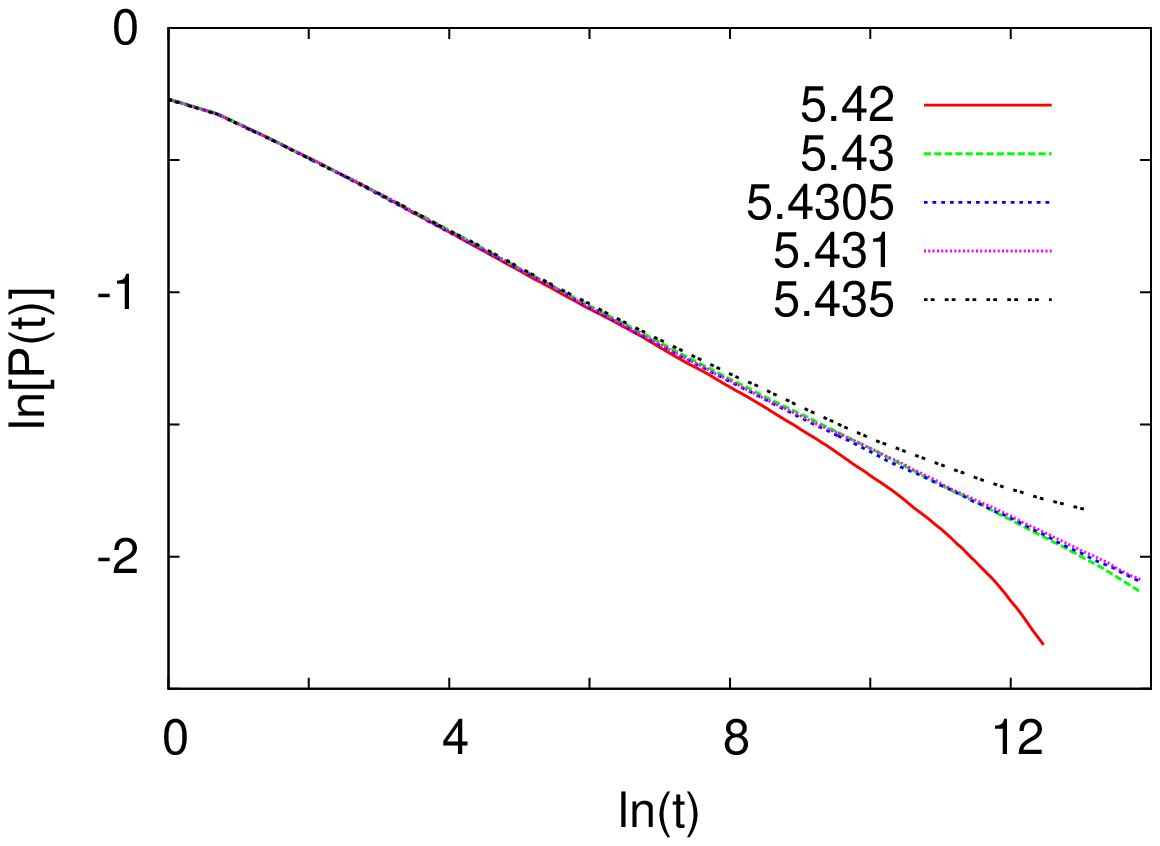}
\includegraphics[width=0.5\linewidth]{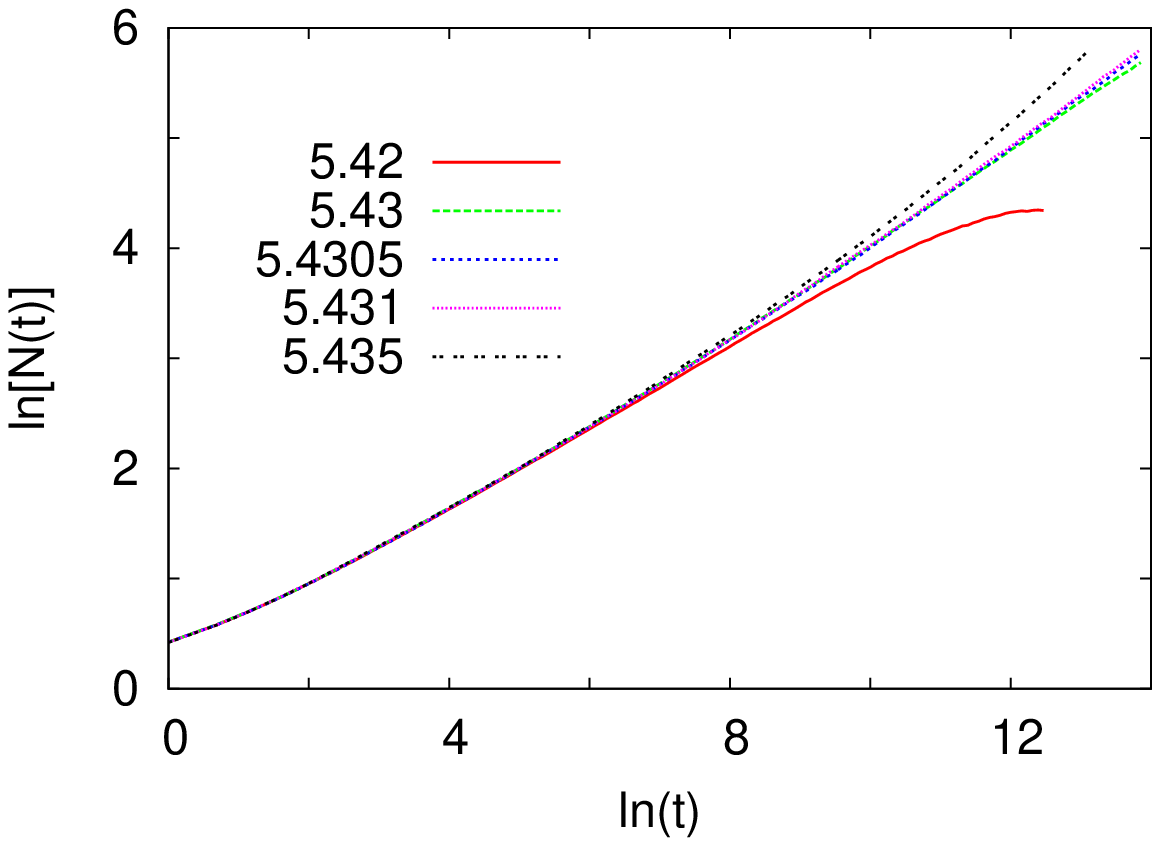}
\caption{\label{np_av}
Dependence of the average survival probability (left) and the average number of individuals (right) on time in the totally asymmetric model with $w=0.4$, for different values of the control parameter $\lambda$. 
}
\end{figure}

\begin{figure}[h]
\includegraphics[width=0.5\linewidth]{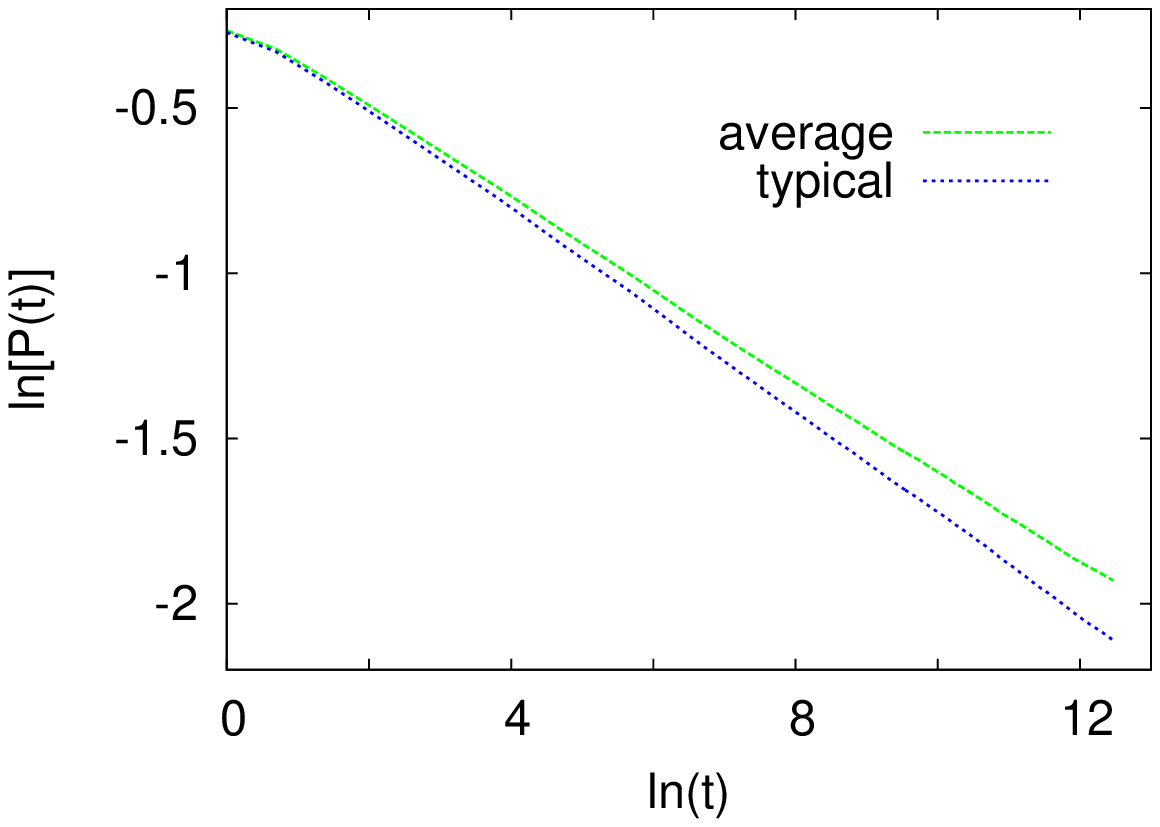}
\includegraphics[width=0.5\linewidth]{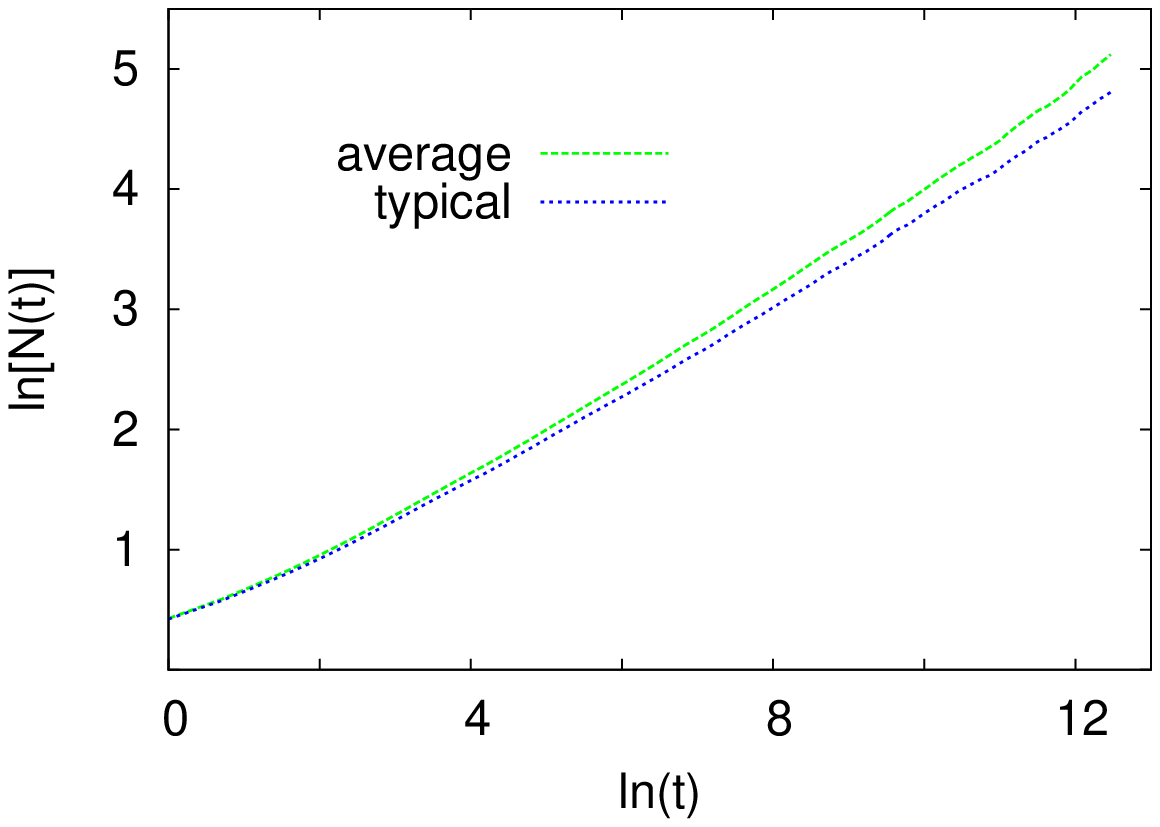}
\caption{
\label{np_typ}
Dependence of the average and typical survival probability (left) and the average and typical number of individuals (right) on time in the totally asymmetric model with $w=0.4$ in the estimated critical point $\lambda_c=5.4305(10)$. 
Linear fits to the data provide the estimates $\delta=0.131(4)$, 
$\delta_{\rm typ}=0.152(5)$, $\eta=0.45(2)$, $\eta_{\rm typ}=0.40(2)$. 
}
\end{figure}

\begin{figure}[h]
\includegraphics[width=0.5\linewidth]{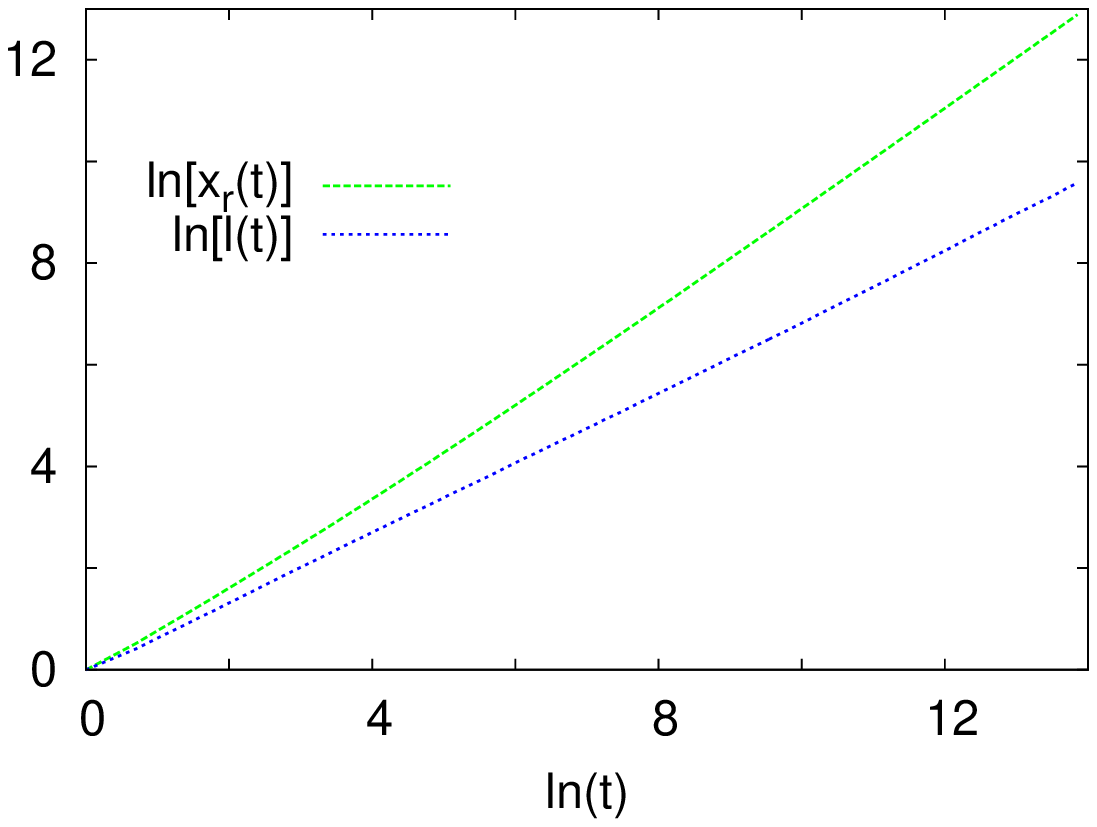}
\includegraphics[width=0.5\linewidth]{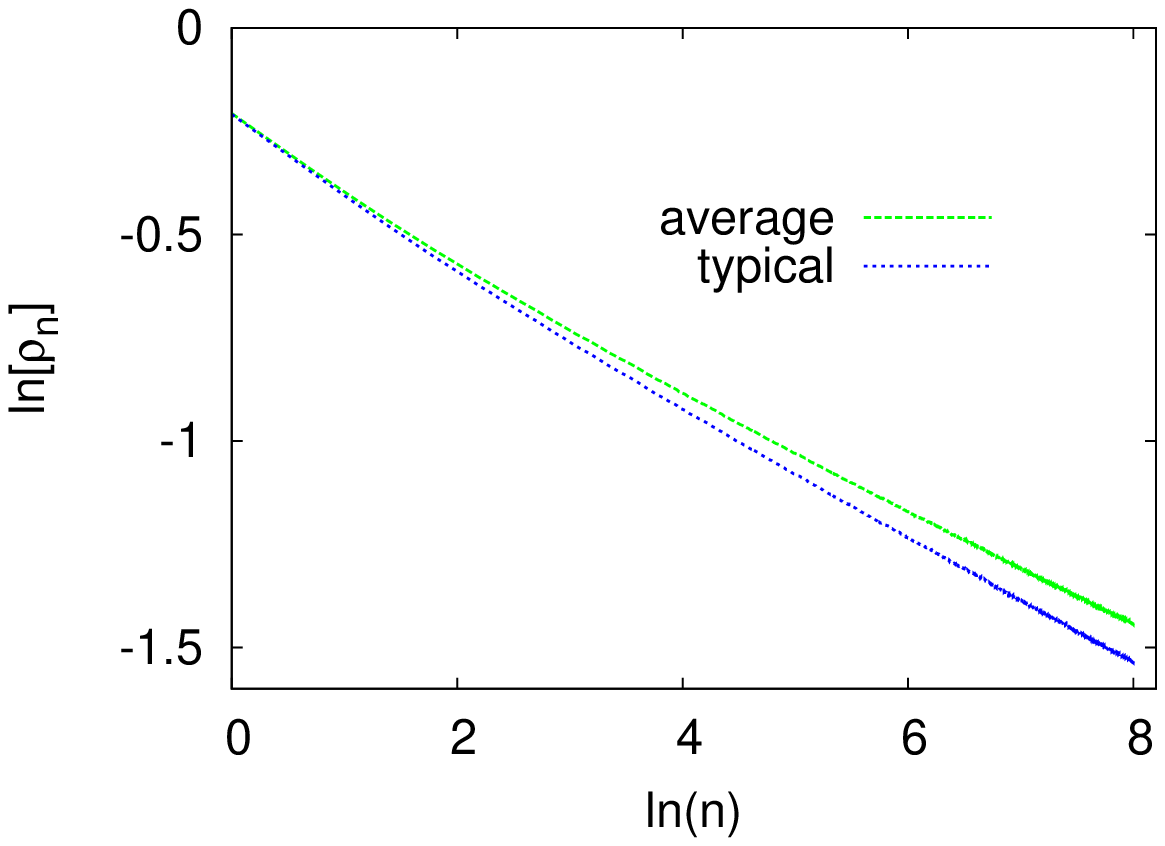}
\caption{
\label{xfl}
Left. Dependence of the average position of the front $\overline{x_r}(t)$ and average spatial extension $\overline{l}(t)$ of the population  on time in the totally asymmetric model with $w=0.4$ in the estimated critical point $\lambda_c=5.4305(10)$. Linear fits to the data give a slope $0.99(1)$ for the $\overline{x_r}(t)$ and $1/z=0.71(1)$ for $\overline{l}(t)$.  
Right.
The average and typical stationary density profiles in the totally asymmetric model with $w=0.4$ in the estimated critical point $\lambda_c=5.4305(10)$ in case of an active boundary at site $0$.   
Linear fits to the data give the estimates $\delta=0.135(5)$, $\delta_{\rm typ}=0.150(5)$.  
}
\end{figure}

The time dependence of the average and typical survival probability, 
$\overline{P}(t)$, $P_{\rm typ}(t)$, the number of individuals, 
$\overline{N}(t)$, $N_{\rm typ}(t)$, shown in Figs \ref{np_av} and \ref{np_typ},
indicate an absorbing phase transition at $\lambda_c=5.4305(10)$. 
%%%%%%%%%%%%%%%%%%%%%%%%%%%%%%%%%%%%%%%%%%%%%%%%%%%%%%%%%%%%%%%%%%%%%%%%%
%%%%%%%%%%%%%%%%%%%%%%%%%%%%%%%%%%%%%%%%%%%%%%%%%%%%%%%%%%%%%%%%%%%%%%%%%
%%%%%%%%%%%%%%%%%%%%%%%%%%%%%%%%%%%%%%%%%%%%%%%%%%%%%%%%%%%%%%%%%%%%%%%%%
%%%%%%%%%%%%%%%%%%%%%%%%%%%%%%%%%%%%%%%%%%%%%%%%%%%%%%%%%%%%%%%%%%%%%%%%%
As can be seen, typical observables vary algebraically in time in the critical point 
and the averages seem to follow a power-law, as well, but it will turn out that 
the averages $\overline{P}(t)$, $\overline{N}(t)$ must have slow, presumably logarithmic corrections, so that the effective exponents seen here must be very different from their true asymptotic values. 
The average position of the front $\overline{x_r}(t)$ of the population advances linearly in the critical point, as expected and the average extension of the population grows as a power of the time, $\overline{l}(t)\sim t^{1/z}$,  see Fig. \ref{xfl}. 
%%%%%%%%%%%%%%%%%%%%%%%%%%%%%%%%%%%%%%%%%%%%%%%%%%%%%%%%%%%%%%%%%%%%%%%%%
%%%%%%%%%%%%%%%%%%%%%%%%%%%%%%%%%%%%%%%%%%%%%%%%%%%%%%%%%%%%%%%%%%%%%%%%%
The average and typical density profiles $\overline{\rho_n}$ and
$\rho_n^{\rm typ}$ are plotted against the distance $n$ from the origin in Fig. 
\ref{xfl}. 
As can be seen, the observables decay as a power of the distance $n$, but, again the averages have slow corrections, which hide the true asymptotic behavior at the time-scales available in the simulations.  
The typical density profile $\rho_n^{\rm typ}$ is found to decay with an exponent $0.150(5)$ which agrees within the error of the measurement with $\delta_{\rm typ}=0.152(5)$. This property, which is valid for the homogeneous, asymmetric model, see Eq. (\ref{Qnb}), thus holds also for the disordered one for transitions of type I, as a consequence of 
the constant velocity of the front. 
We mention that the effective exponents of average quantities 
$\overline{P}(t)$,  $\overline{Q_n}$, and $\overline{\rho_n}$ also agree with each other within the errors of measurements, and have the value $\delta_{\rm eff}=0.131(4)$ for $w=0.4$. 

\begin{figure}[h]
\includegraphics[width=0.7\linewidth]{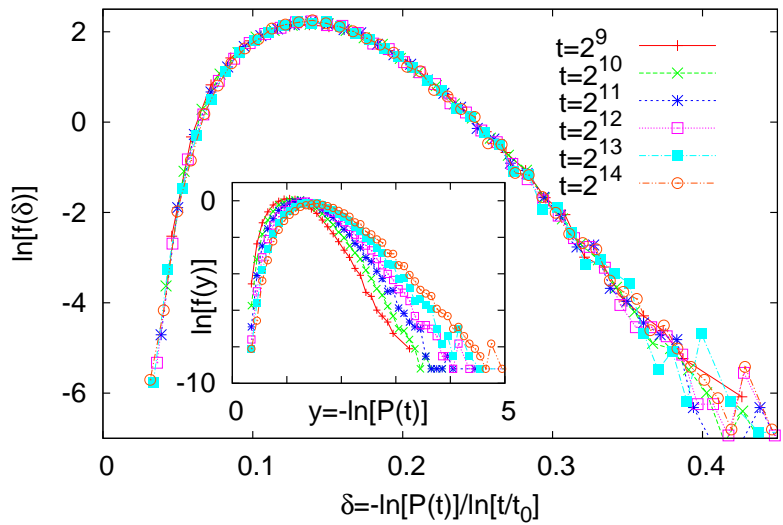}
\caption{
\label{ptd.4}
Scaling plot of the histogram of the survival probability in the estimated critical point at different times. An optimal collapse is obtained by the parameter $t_0=0.25$. The inset shows the unscaled data. 
}
\end{figure}

We can observe in the figures that the average and the typical quantities are characterized by different critical exponents (the former by effective ones) in general. In other words, they are non-self-averaging, which is a well-known feature of disordered systems \cite{wd,bouchaud}.
In order to have more insight into this phenomenon, 
we have constructed the histogram of the logarithm of $P(t,\Omega)$ 
from data obtained in $10^5$ independent random environments. 
As can be seen in Fig \ref{ptd.4}, the distributions are broadening with increasing time and, plotting the distribution $f(\delta)$ of the scaling variable 
\be 
\delta=-\ln[P(t)]/\ln(t/t_0)
\ee
with a time-scale $t_0$ (which is irrelevant in the limit of long times), 
we obtain a good scaling collapse. 
%%%%%%%%%%%%%%%%%%%%%%%%%%%%%%%%%%%%%%%%%%%%%%%%%%%%%%%%%%%%%%%%%%%%%%%%%
%%%%%%%%%%%%%%%%%%%%%%%%%%%%%%%%%%%%%%%%%%%%%%%%%%%%%%%%%%%%%%%%%%%%%%%%%
The scaling variable $\delta$ is nothing but a sample-dependent 
effective decay exponent of the survival probability. 
As opposed to the homogeneous system, which is characterized by one single value of $\delta$, the disordered model has a broad distribution of effective decay exponents with a probability density $f(\delta)$. 
This kind of behavior, termed as {\it multiscaling}, appears also in the autocorrelations of disordered critical quantum spin chains \cite{ky,ijr}. 
The exponent $\delta_{\rm typ}$ describing the decay of typical survival probability as $P_{\rm typ}(t)\sim t^{-\delta_{\rm typ}}$ is the first moment of $f(\delta)$: 
\be 
\delta_{\rm typ}=\int_0^{\infty}\delta f(\delta)d\delta,
\ee
whereas the average survival probability $\overline{P}(t)=\int_0^{\infty}e^{-\delta\ln t}f(\delta)d\delta$ is determined by $f(\delta)$ around its lower edge 
$\delta_0\equiv \sup\{\delta : f(\delta)=0\}$. 
Assuming that $f(\delta)=(\delta-\delta_0)^a$ in leading order as $\delta\to\delta_0$, it is easy to show that 
\be 
\overline{P}(t)\sim t^{-\delta_0}(\ln t)^{-(a+1)} 
\ee
for large $t$. 
So, if $\delta_0>0$ the decay is algebraic with a multiplicative logarithmic factor, which results in a considerable shift of the effective exponent for finite times.  Otherwise the time-dependence is purely logarithmic, which is slower than any power of $t$. 
It is hard to judge from the numerical data whether $\delta_0$ is positive or not, especially for small $w$. Plotting $\ln f(\delta)$ against $\ln\delta$ using histograms of higher resolution (not shown) one can see that the slope is not constant but increases steadily for decreasing $\delta$, which is in favor of a positive (but small) $\delta_0$.  
As can be seen, the effective decay exponent $\delta_{\rm eff}=0.131(4)$ obtained from the numerical data is deeply in the domain where $f(\delta)>0$, thus it must not be the asymptotic value. 

\begin{figure}[h]
\includegraphics[width=0.7\linewidth]{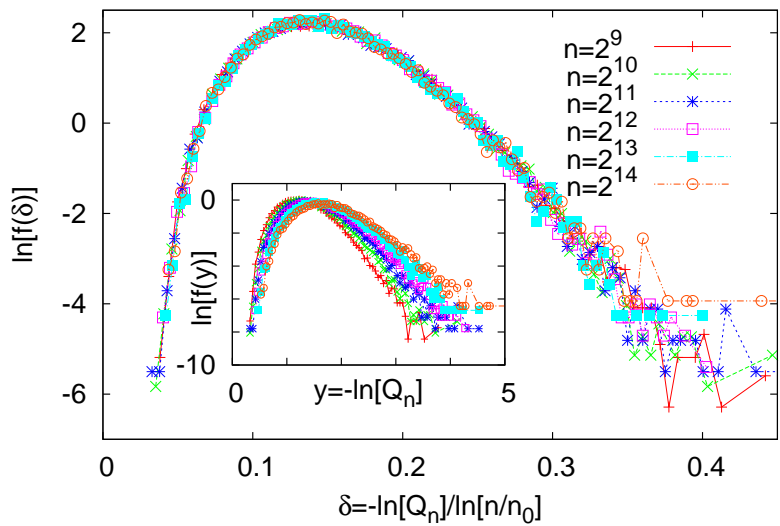}
\caption{
\label{pxd}
Scaling plot of the histogram of the hitting probability at different distances from the origin in the estimated critical point. An optimal collapse is obtained by the parameter $n_0=0.1$. The inset shows the unscaled data. 
}
\end{figure}

We have also constructed histograms of the hitting probability $Q_n(\Omega)$ at different distances from the origin. These have the same multiscaling property as $P(t)$, see Fig. \ref{pxd}. 
%%%%%%%%%%%%%%%%%%%%%%%%%%%%%%%%%%%%%%%%%%%%%%%%%%%%%%%%%%%%%%%%%%%%%%%%%
%%%%%%%%%%%%%%%%%%%%%%%%%%%%%%%%%%%%%%%%%%%%%%%%%%%%%%%%%%%%%%%%%%%%%%%%%

\begin{figure}[h]
\includegraphics[width=0.7\linewidth]{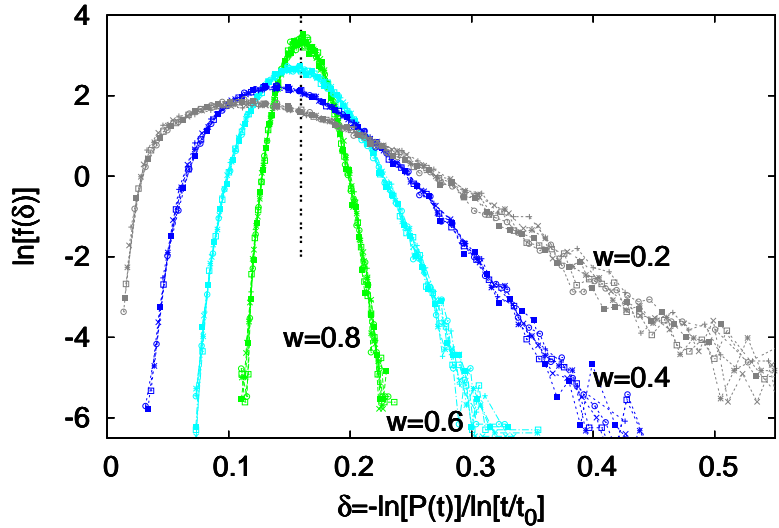}
\caption{
\label{ptdcommon}
Spectrums of the decay exponent of the survival probability in the totally asymmetric model for different values of the strength of disorder $w$. 
The vertical line indicates $\delta$ of the homogeneous model ($w=1$). 
}
\end{figure}

The spectrums $f(\delta)$ of the decay exponent of the survival probability for different values of the strength of disorder $w$ are shown in Fig. \ref{ptdcommon}. 
As can be seen, they are broader for stronger disorder (smaller $w$), and seem to have a finite width for any finite disorder, in accordance with the prediction of the Harris criterion. The most probable value of the distribution, as well as the first moment ($\delta_{\rm typ}$) is decreasing with decreasing $w$, see the data in Table \ref{table_w}. 
%%%%%%%%%%%%%%%%%%%%%%%%%%%%%%%%%%%%%%%%%%%%%%%%%%%%%%%%%%%%%%%%%%%%%%%%%
%%%%%%%%%%%%%%%%%%%%%%%%%%%%%%%%%%%%%%%%%%%%%%%%%%%%%%%%%%%%%%%%%%%%%%%%%
Other exponents vary continuously with $w$, as well; $\eta_{\rm typ}$ is increasing while $z$ is decreasing with decreasing $w$. 
%%%%%%%%%%%%%%%%%%%%%%%%%%%%%%%%%%%%%%%%%%%%%%%%%%%%%%%%%%%%%%%%%%%
\begin{table}[h]
\begin{center}
\begin{tabular}{|l|l|l|l|l|}
\hline  $w$ & $\lambda_c$ & $\delta_{\rm typ}$ & $\eta_{\rm typ}$ & $z$   \\
\hline
\hline  0.8 & 3.7053(3)  & 0.160(2) & 0.32(1) & 1.56(2) \\
\hline  0.6 & 4.3196(4)  & 0.156(4) & 0.34(1) & 1.52(2) \\
\hline  0.4 & 5.4305(10) & 0.152(5) & 0.40(1) & 1.39(2) \\
\hline  0.2 & 8.305(10)  & 0.13(1)  & 0.55(2) & 1.28(3) \\
 \hline
\end{tabular}
\end{center}
\caption{\label{table_w} 
The location $\lambda_c$ of the critical point and estimated critical exponents of the totally asymmetric model for different values of the strength of the disorder.}
\end{table}
%%%%%%%%%%%%%%%%%%%%%%%%%%%%%%%%%%%%%%%%%%%%%%%%%%%%%%%%%%%%%%%%%%%%%%

\section{Numerical results for the partially asymmetric model}
\label{pascp}

We have performed numerical simulations of the partially asymmetric model, as well,  mainly for $a=2/5$. We have checked by simulations of the homogeneous model that this value of the asymmetry parameter is above $a^*(w)$, 
so that the phase transition is of type II. 

\begin{figure}[h]
\includegraphics[width=0.5\linewidth]{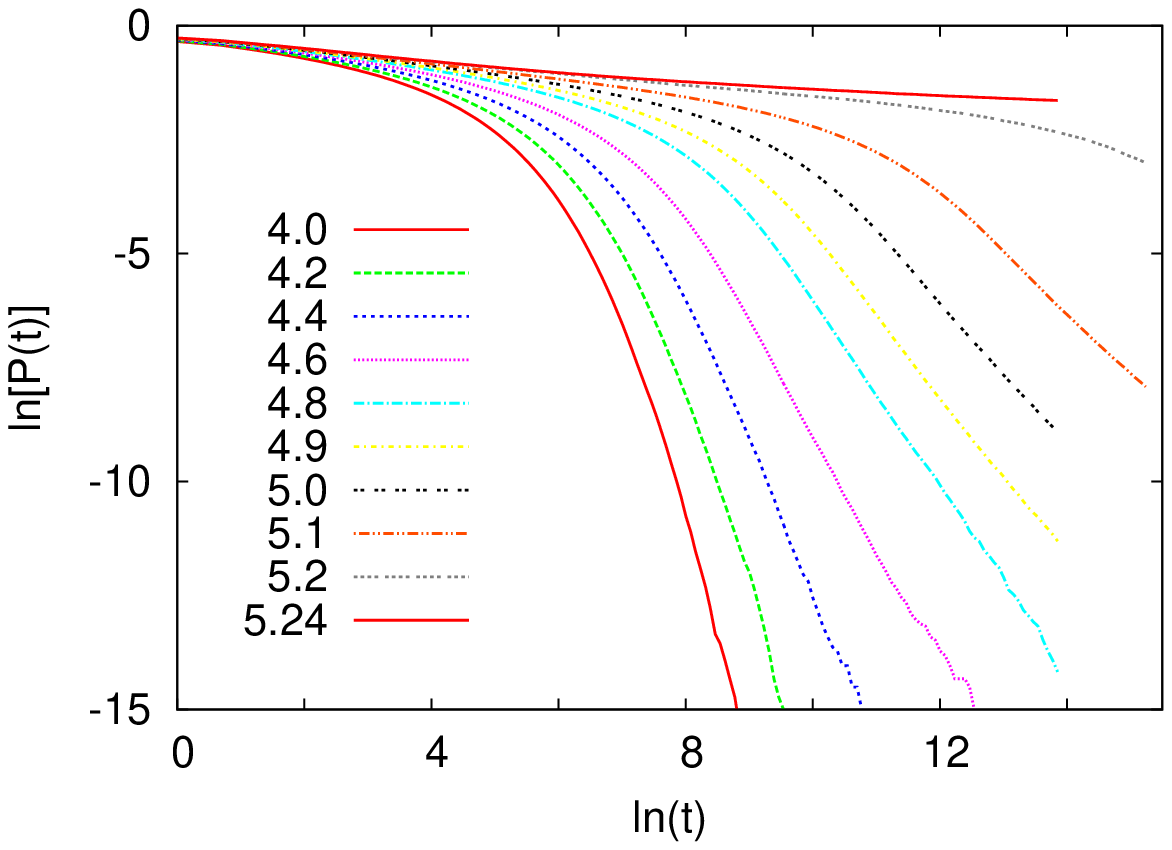}
\includegraphics[width=0.5\linewidth]{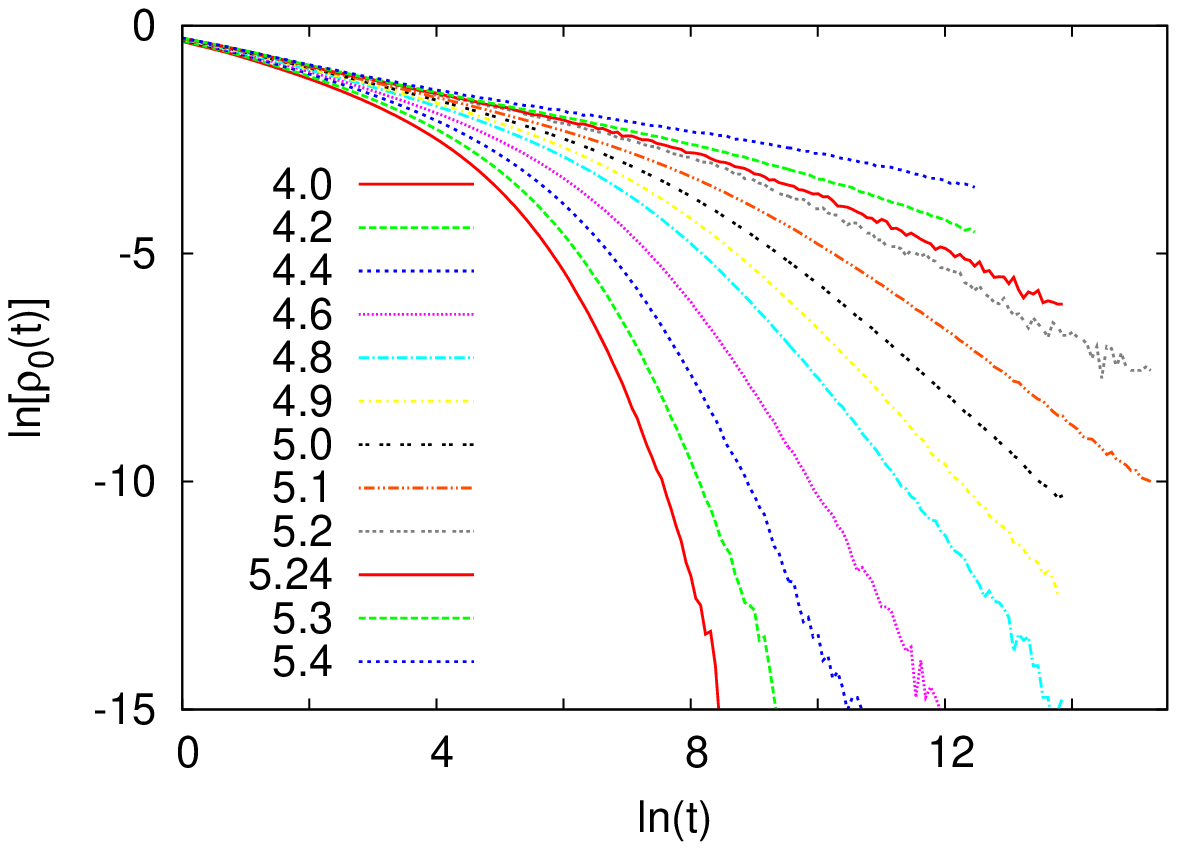}
\caption{
\label{gp.4}
Left: The logarithm of the average survival probability  plotted against the 
logarithm of time in the model with $a=2/5$ for different values of the control parameter $\lambda$. 
Right. The logarithm of the density at the origin as a function of the logarithm of time in the same model. 
}
\end{figure}

As it has been predicted, below the critical point, a Griffiths phase appears, where the average survival probability decays algebraically with the time with exponents depending on $\lambda$, see Fig. \ref{gp.4}. 
The density $\rho_0(t)$ at the origin shows a similar behavior, 
see the right panel of the same figure.  
%%%%%%%%%%%%%%%%%%%%%%%%%%%%%%%%%%%%%%%%%%%%%%%%%%%%%%%%%%%%%%%%%%%%%%%%%
%%%%%%%%%%%%%%%%%%%%%%%%%%%%%%%%%%%%%%%%%%%%%%%%%%%%%%%%%%%%%%%%%%%%%%%%%

\begin{figure}[h]
\includegraphics[width=0.5\linewidth]{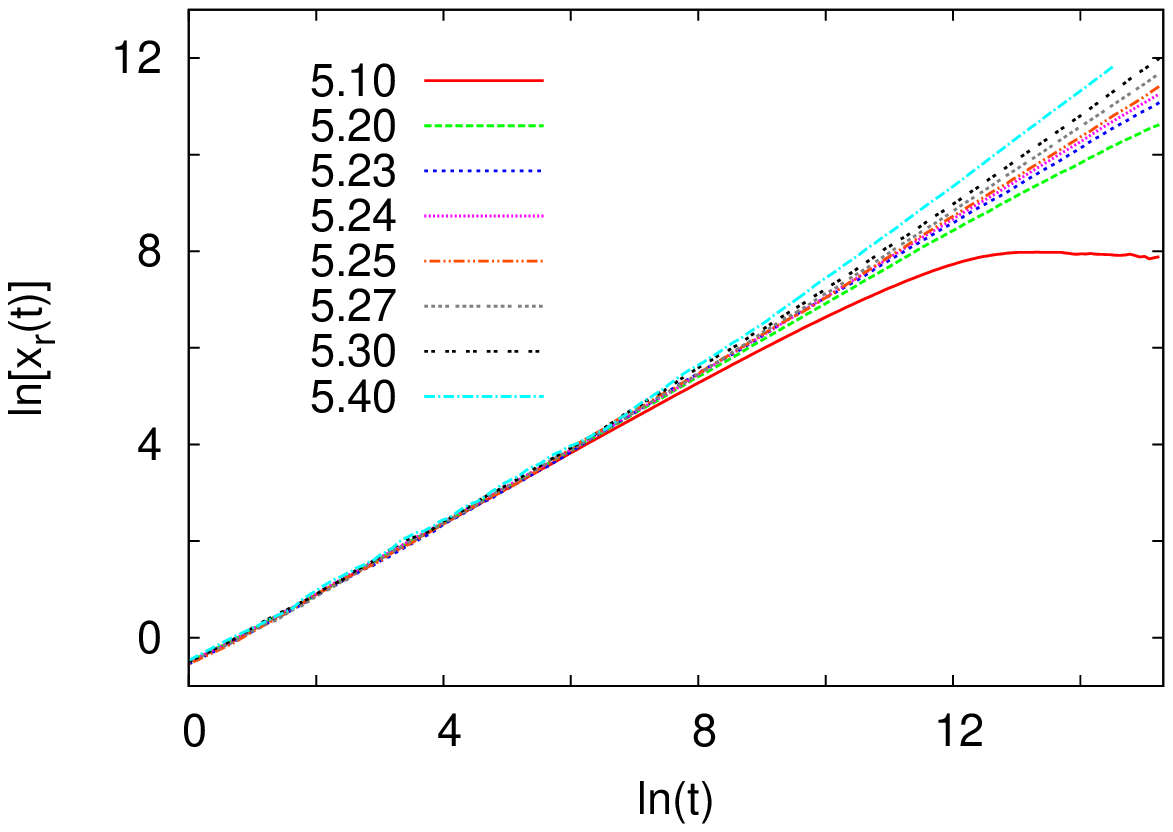}
\includegraphics[width=0.5\linewidth]{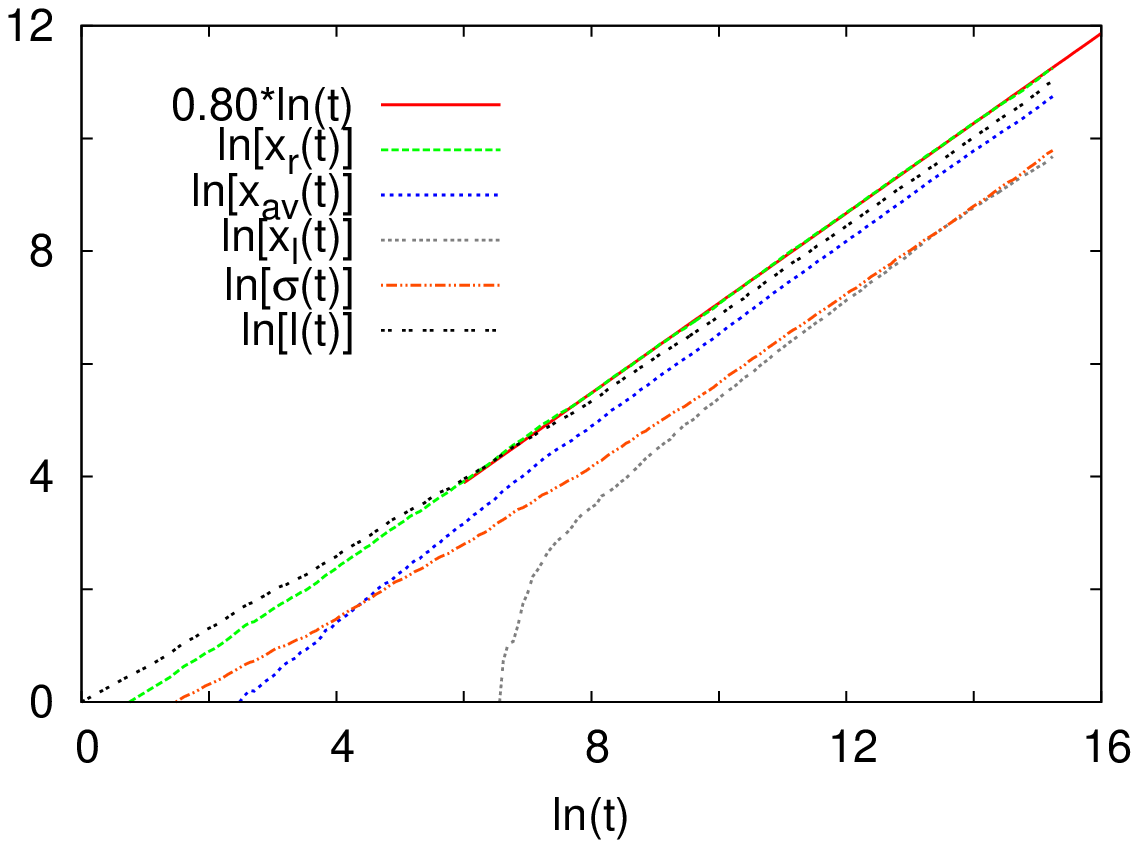}
\caption{
\label{front}
Left. Dependence of the average position $\overline{x_r}(t)$ of the front on time  
in the model with asymmetry parameter $a=2/5$ for different values of the control parameter. 
Right. 
Dependence of different characteristics of the location of the population on time at $\lambda=5.24$. $x_{\rm av}(t)$ and $\sigma(t)$ denote the average value and the standard deviation of the position of individuals at time $t$.
}
\end{figure}

An important question to be clarified is how the population shifts 
as time elapses. 
The dependence of the position of the front on time is shown 
in Fig. \ref{front}. 
In the one-dimensional symmetric model ($a=1/2$) it is known that another Griffiths phase exists above the critical point, where the front spreads anomalously as $\overline{x_r}(t)\sim t^{1/z_r}$ with a $\lambda$-dependent diffusion exponent 
$1/z_r(\lambda)<1$ \cite{bds}. As $\lambda\to\lambda_c$, the diffusion exponent 
tends to zero and, in the critical point, the spreading becomes ultra-slow,
i.e. $\overline{x_r}(t)\sim (\ln t)^2$. 
A similar Griffiths phase can be observed in the asymmetric model, as well, 
where, by decreasing $\lambda$, the diffusion exponent $1/z_r$ decreases starting from $1$. But, rather than tending to zero, $1/z_r$ seems to remain finite 
($1/z_r=0.81(3)$) when the critical point is reached at $\lambda_c=5.24(2)$.
%%%%%%%%%%%%%%%%%%%%%%%%%%%%%%%%%%%%%%%%%%%%%%%%%%%%%%%%%%%%%%%%%%%%%%%%%
%%%%%%%%%%%%%%%%%%%%%%%%%%%%%%%%%%%%%%%%%%%%%%%%%%%%%%%%%%%%%%%%%%%%%%%%%
Another difference with respect to the symmetric model is that also the left end of the population shifts rightwards for long times i.e. $\overline{x_l}(t)\to\infty$ (as far as the model is below the upper transition point). 
The corresponding diffusion exponent above the extinction transition (and below the upper transition point) can be shown to be related to the exponent $1/z_l$ 
appearing in the decay of the average density at the origin,
\be
\rho_0(t)\sim t^{-1/z_l},
\label{rho0}
\ee
as follows. 
As mentioned in section \ref{model}, this anomalous decay is caused 
by locally supercritical domains, whose lifetime distribution must have, 
according to Eq. (\ref{rho0}), the long time asymptotics $P_>(\tau)\sim \tau^{-1/z_l}$.  
The traveling time $t$ of the left end of the population from the origin to a far away site $n$ can then be written as a sum of $O(n)$ waiting times $\tau_i$, 
\be
t\sim \sum_i^{O(n)}\tau_i.
\label{travel}
\ee 
Such a sum is dominated by the maximal term whenever $z_l>1$, see e.g. Ref. \cite{bouchaud}, and, applying extreme value statistics, 
we obtain $t\sim n^{z_l}$. Thus the left end of the population advances for long times as 
\be
\overline{x_l}(t)\sim t^{1/z_l}.
\ee 
The numerical data are satisfactorily consistent with this relation though, for a clearer observation of the asymptotics, longer times would be needed. 

The drift of the population is, however, not necessarily sublinear even in the case of a transition of type II.   
An alternative possibility is that, increasing the control parameter in the subcritical Griffiths phase, $z_l$ is increasing but at the transition point still $z_l<1$. 
In that case, the sum in Eq. (\ref{travel}) is proportional to $n$ and, consequently, both the left end and the front move with a constant velocity.  
But, even in this case, a Griffiths phase appears below the transition point (but not above it). 
This scenario is expected to be realized for a weak asymmetry (but still  
for  $a>a^*(w)$). 

\begin{figure}[h]
\includegraphics[width=0.5\linewidth]{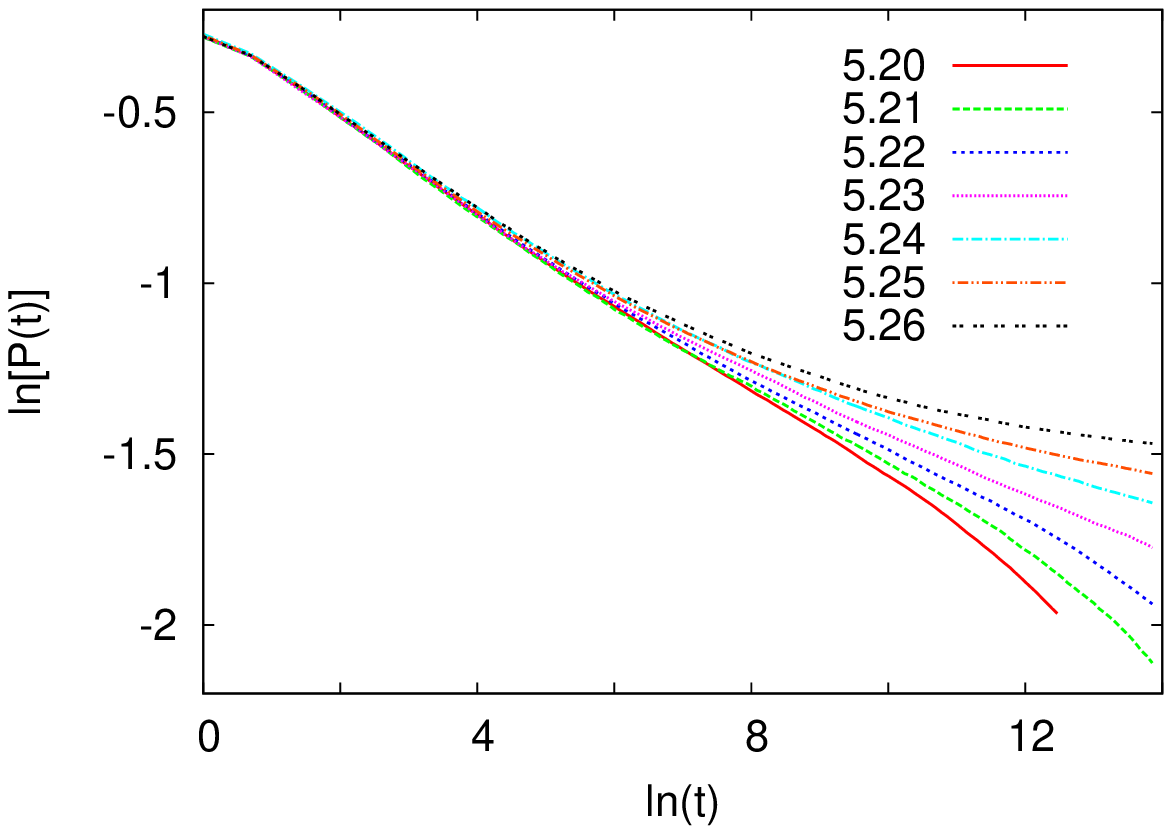}
\includegraphics[width=0.5\linewidth]{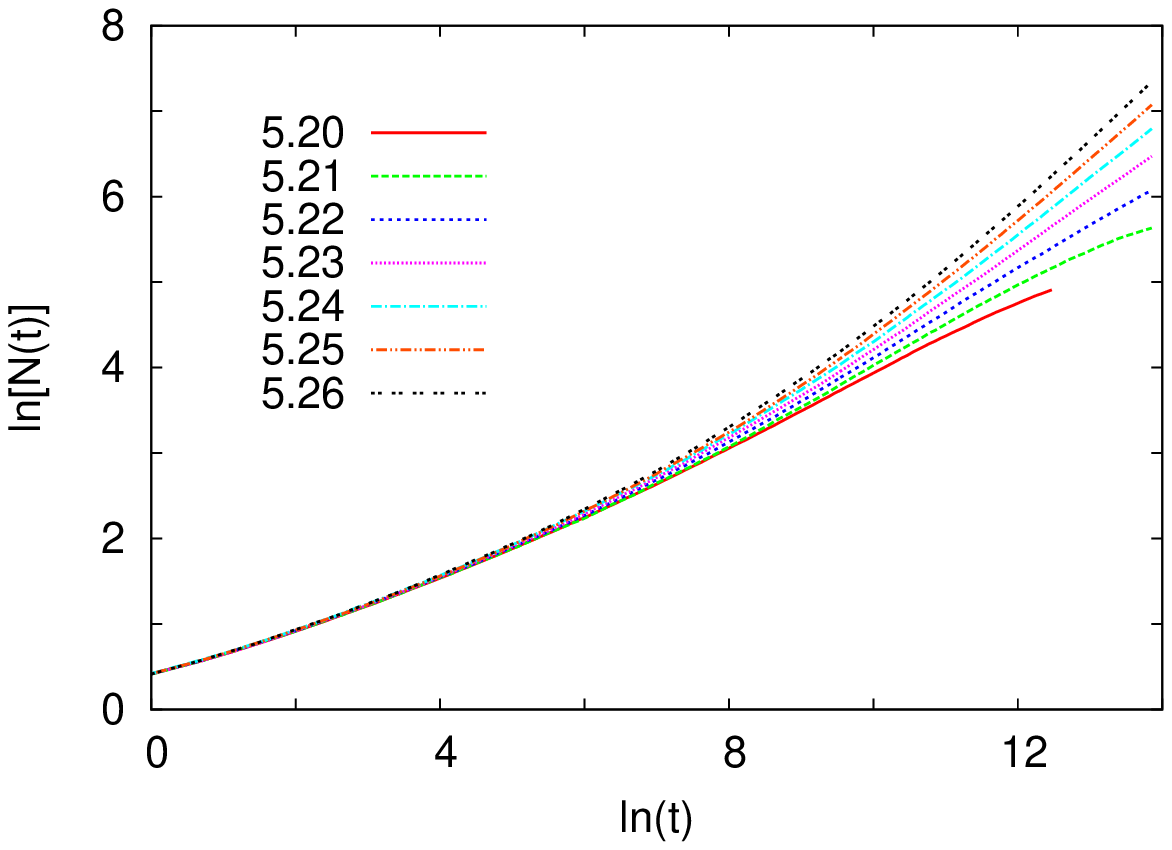}
\caption{
\label{np.4}
Left. The logarithm of the average survival probability  plotted against the 
logarithm of the time in the model with $a=2/5$ for different values of the control parameter. 
Right. The logarithm of the average number of particles as a function of the logarithm of the time in the same model. 
}
\end{figure}

Fig. \ref{np.4} shows the dependence of the average survival probability and the average number of individuals on time in the vicinity of the critical point.  
%%%%%%%%%%%%%%%%%%%%%%%%%%%%%%%%%%%%%%%%%%%%%%%%%%%%%%%%%%%%%%%%%%%%%%%%%
%%%%%%%%%%%%%%%%%%%%%%%%%%%%%%%%%%%%%%%%%%%%%%%%%%%%%%%%%%%%%%%%%%%%%%%%%
One can observe that, for smaller values of $\lambda$, the curvature of 
$\ln\overline{P}(t)$ as a function of $\ln t$ becomes negative for long times, 
indicating that these values of $\lambda$ are in the extinction phase. 
For shorter times the curvature is positive and the inflection point is shifted 
toward longer times as $\lambda$ is increased in the extinction phase and, ultimately, it goes to infinity in the critical point. 
It is, however, hard to judge from the numerical data where the accurate location of the critical point is and whether the slope tends to a non-zero constant there for long times (which corresponds to a power-law decay) or tends to zero, which would be the case, for instance, for a logarithmic decay given in Eq. (\ref{actP}). 
As can be seen in Fig. \ref{np.4}, the average number of particles seems to increase algebraically with the time but with strong corrections and, in accordance with the tendency observed in the totally asymmetric model,  the (effective) exponent $\eta$ exceeds the value of the homogeneous model.  

After having seen the main features of the critical system, let us return to the problem of determining the location of the critical point.  
A usual method for the symmetric, disordered contact process is to plot $\ln[\overline{P}(t)]$ against $\ln[\overline{N}(t)]$ and to look for $\lambda$ at which the curve is, according to Eqs. (\ref{actP}) and (\ref{actN}),  asymptotically linear with a slope $-\tilde\eta/\tilde\delta$ \cite{vd}. This works equally well for the homogeneous model, where the slope is $-\eta/\delta$.  
We have seen that, for the disordered, asymmetric model,  $\delta$ is very 
small or may even vanish, while $\eta$ is relatively large, thus their ratio is very large, making the above method inaccurate. 
Therefore, we need another indicator of the critical point. 
A possible candidate is the ratio of the average hitting probability 
and the average survival probability at time $t=n$, 
\be  
I(n)\equiv\overline{Q_n}/\overline{P}(t=n).
\ee
In the weak-survival phase, 
$\lim_{n\to\infty}\overline{Q_n}=\lim_{t\to\infty}\overline{P}(t)>0$, therefore 
$I(n)$ tends to $1$ as $n\to\infty$. 
As opposed to this, in the extinction phase, $\overline{P}(t)$ decays as a power of $t$, while, as the population becomes localized in space after an initial shift, $\overline{Q_n}$ decays faster than any power of $n$. 
Hence $I(n)$ goes to zero rapidly here. 
At the critical point, let us assume an algebraic displacement 
$\overline{x_r}(t)\sim t^{1/z_r}$ of the front, as seen in the simulations. 
Then, using the relation 
\be 
\overline{P}(t)\sim \overline{Q}_{n=\overline{x_r}(t)},
\ee
$I(n)$ can be shown to have different asymptotics depending on whether 
$\overline{P}(t)\sim t^{-\delta}$ or 
$\overline{P}(t)\sim (\ln t)^{-\tilde\delta}$. 
In the former case, it goes to zero as $I(n)\sim n^{-\delta(z_r-1)}$ while, in the latter,
$\overline{Q_n}\sim (z_r\ln n)^{-\tilde\delta}$ and
$I(n)$ tends to a positiv constant, which is less than $1$.  

\begin{figure}[h]
\includegraphics[width=0.5\linewidth]{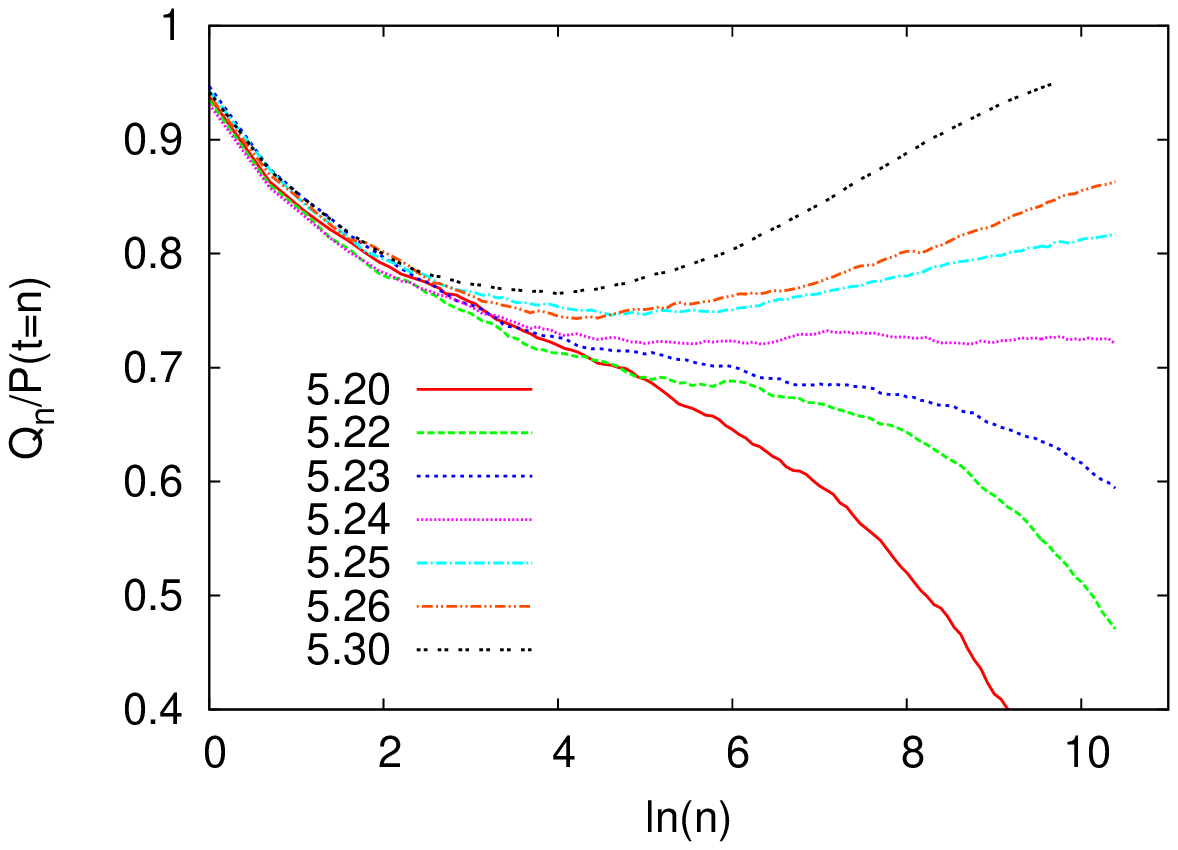}
\includegraphics[width=0.5\linewidth]{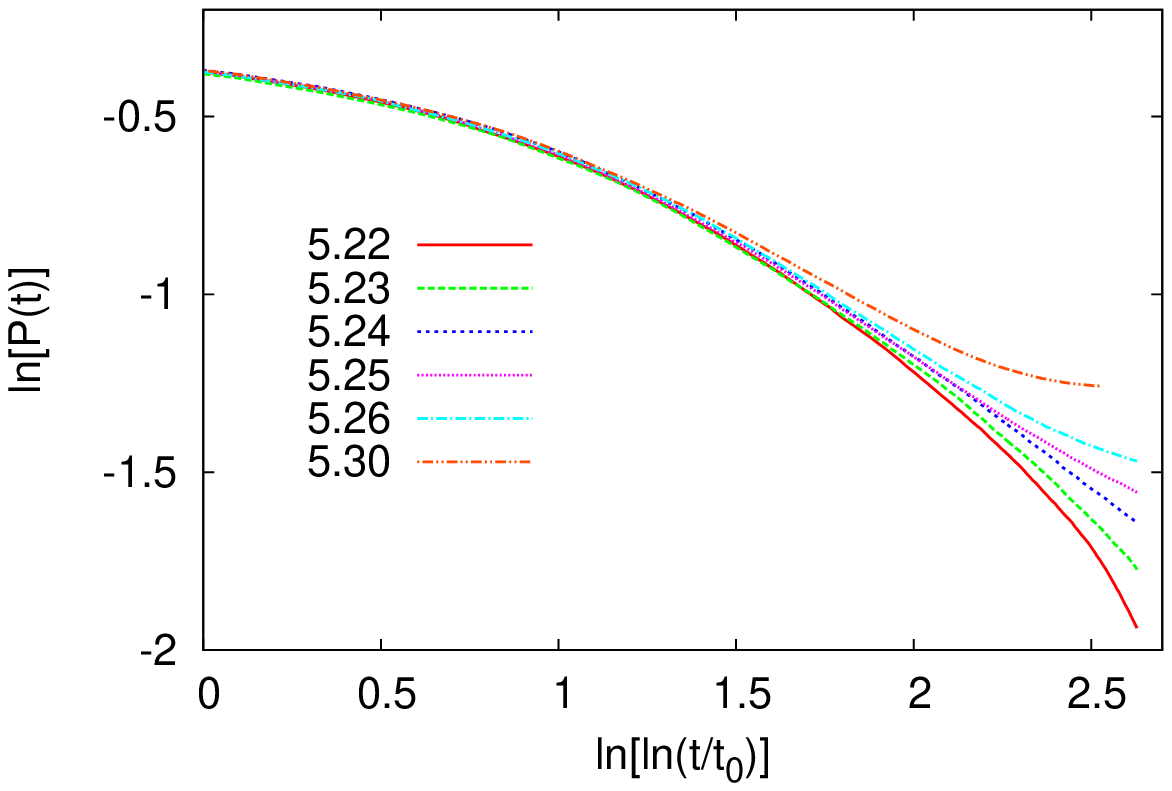}
\caption{
\label{act.4}
Left. The ratio $\overline{Q}_n/\overline{P}(t=n)$ plotted against $\ln(n)$
in the model with $a=2/5$ for different values of the control parameter in the vicinity of the critical point. $\overline{Q}_n$ has been obtained by measurements in $10^4$ random environments. 
Right. The logarithm of the average survival probability as a function of 
$\ln[\ln(t/t_0)]$ with $t_0=1$. 
}
\end{figure}

Numerical results for $I(n)$ are shown in Fig. \ref{act.4}. 
As can be seen, $I(n)$ tends to a constant at $\lambda=5.24$, while for larger (smaller) values of $\lambda$ it increases (decreases) with $n$. 
%%%%%%%%%%%%%%%%%%%%%%%%%%%%%%%%%%%%%%%%%%%%%%%%%%%%%%%%%%%%%%%%%%%%%%%%%
%%%%%%%%%%%%%%%%%%%%%%%%%%%%%%%%%%%%%%%%%%%%%%%%%%%%%%%%%%%%%%%%%%%%%%%%%
This behavior is compatible with an activated scaling $\overline{P}(t)\sim (\ln t)^{-\tilde\delta}$ of the average survival probability (and that of the average hitting probability) at least at the time scales available by the simulations. 
However, for an accurate estimation of $\tilde\delta$, for example, by plotting 
$\ln[\overline{P}(t)]$ against $\ln[\ln(t)]$ as it has been done in the right panel of Fig. \ref{act.4}, much longer simulation times would be needed. 

\begin{figure}[h]
\includegraphics[width=0.7\linewidth]{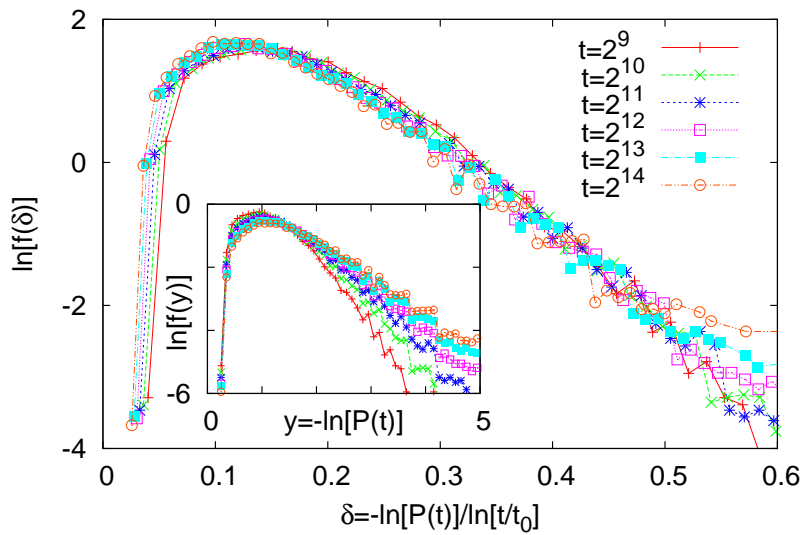}
\caption{
\label{ptda.4}
The spectrum of effective exponents of the survival probability in the model with $a=2/5$ in the estimated critical point for different times. The time scale is set to $t_0=1$. The inset shows the histograms of the logarithm of the survival probability. 
}
\end{figure}

We have also investigated the spectrum of the effective decay 
exponents $\delta=-\ln[P(t)/\ln(t/t_0)]$ in the estimated critical point, 
see Fig. \ref{ptda.4}.
In contrast with the totally asymmetric model, the distributions are not invariant in time but slowly shift toward zero with increasing time, and this shift cannot be eliminated by an appropriate choice of $t_0$.  
This picture is qualitatively similar to that can be seen for the symmetric model (not shown), where the average survival probability obeys the activated scaling given Eq. (\ref{actP}).
But again, to infer the true asymptotics of the distribution $f(\delta)$, the time scales available by the simulations are not sufficiently large. 
%%%%%%%%%%%%%%%%%%%%%%%%%%%%%%%%%%%%%%%%%%%%%%%%%%%%%%%%%%%%%%%%%%%%%%%%%
%%%%%%%%%%%%%%%%%%%%%%%%%%%%%%%%%%%%%%%%%%%%%%%%%%%%%%%%%%%%%%%%%%%%%%%%%

\section{Discussion}
\label{discussion}

In this work, we have studied the dynamics of a population in a convective, random environment by an asymmetric variant of the one-dimensional contact process in which creations of new individuals occurs with larger rates 
in either direction. 
We have seen that spatial disorder is a relevant perturbation in the asymmetric model and the resulting behavior of the model is much different from that of the homogeneous one. 
At the absorbing phase transition, the different dynamical quantities are non-self-averaging. The survival probability shows multiscaling, i.e. it is characterized by a broad spectrum of decay exponents.   
As the population is forced to drift across a quenched random environment, 
its growth is affected in fact by an effective, time-dependent environment, 
which is steadily moving relative to the population. 
It is therefore not surprising that the model has different properties from that of its static variant, the disordered, symmetric contact process.   
The most apparent difference is that the front of the population advances as a power of the time in the extinction transition point
rather than logarithmically and the corresponding diffusion exponent $1/z_r$
is influenced by the asymmetry parameter. 
For strong enough asymmetry, the motion of the population is linear in time; 
otherwise, the diffusion exponent is less than one and decreases with 
weakening asymmetry. As $a\to 1/2$, it is expected to vanish, $1/z_r\to 0$, and, at $a=1/2$, the power law is replaced by a logarithmic dependence on time. 
This behavior is reminiscent of the dynamics of random walks in a disordered environment, see e.g. Ref. \cite{bouchaud}. 
In that problem, the typical displacement is $|x(t)|\sim (\ln t)^2$ if the 
distributions of jump rates to the left and to the right are identical, 
while, in case of a bias in either direction, it switches to the form $|x(t)|\sim t^{\mu}$, where $\mu\le 1$ is increasing with the bias and $\mu=1$ beyond a certain point.  
Another difference to the infinite-randomness critical behavior of the 
symmetric model is that the average dynamical quantities seem to follow non-universal power-laws (possibly with the exception of the survival probability for weak asymmetry) rather than activated scaling given in Eqs. (\ref{actP}-\ref{actl}).

We have also seen that the disorder does not affect only the properties in the extinction transition point but also around it. 
For a sufficiently weak asymmetry, a Griffiths phase emerges below the extinctions transition, where, although the population gets extinct in the limit $t\to\infty$, it still survives for very long time in certain atypical samples, 
so that the average survival probability decays algebraically. 
Moreover, above the transition point another Griffiths phase appears where the front of the population advances anomalously slowly with a $\lambda$-dependent diffusion exponent. 

It has also turned out that relying exclusively on results of Monte Carlo 
simulations, many features of the model are hard to clarify and several questions are left open such as the asymptotic form of the time dependence of average quantities. 

In this work, we have restricted ourselves to one spatial dimension, which 
may reflect some properties of higher dimensional generalizations of the model, such as the appearance of a Griffiths phase, 
but it is suitable only for the description of very special cases when the population is constrained to live in a quasi-one-dimensional domain, e.g. a river.
It would be desirable to extend the present investigations to a more realistic two-dimensional variant of the model where spreading of the population is more favorable in a given direction than in the other ones. 

Viewed in a wider context, the present study suggests that systems of interacting degrees of freedom that are forced to drift across a disordered environment may have, in general, a special behavior different from that of the corresponding static model. It would be instructive to study this problem in simpler, more easily tractable models.

%%%%%%%%%%%%%%%%%%%%%%%%%%%%%%%%%%%%%%%%%%%%%%%%%%%%%%%%%%%%%%%%%%%%%
%%%%%%%%%%%%%%%%%%%%%%%%%%%%%%%%%%%%%%%%%%%%%%%%%%%%%%%%%%%%%%%%%%%%%
\appendix
\section{}   

In the quantum Hamilton formalism (see e.g. Ref. \cite{schutz}), the state of site $i$ is described by a two dimensional vector $|\eta_i\rangle$, the basis vectors $|0\rangle$ and  $|1\rangle$ corresponding to the empty and occupied states, respectively. 
The state of a system of size $L$  at time $t$ can be expanded in the 
basis $\{|\eta\rangle\}$ where $|\eta\rangle\equiv \otimes_{i=1}^L|\eta_i\rangle$, $\eta_i=0,1$ as $|P(t)\rangle = \sum_{\eta}P_{\eta}(t)|\eta\rangle$. 
The master equation of the process can then be written in the form 
\be 
\partial_t|P(t)\rangle=-H|P(t)\rangle, 
\ee
with the Hamilton operator (also called as Liouville operator) $H$:
\be 
H=-\sum_{i}\mu M_i - \sum_{i}\lambda_iL_i -  \sum_{i}\kappa_{i+1}K_i. 
\ee 
Here, $M_i$ acts non-trivially only on site $i$: 
\be 
M_i= {\bf 1}_1\otimes\dots\otimes {\bf 1}_{i-1}\otimes(s_i^+-n_i)\otimes{\bf 1}_{i+1}\otimes\dots\otimes {\bf 1}_L,
\ee
whereas 
$L_i$ and $K_i$ act non-trivially on sites $i$ and $i+1$:
\beqn 
L_i= {\bf 1}_1\otimes\dots\otimes {\bf 1}_{i-1}\otimes n_i\otimes (s_{i+1}^--v_{i+1})\otimes{\bf 1}_{i+2}\otimes\dots\otimes {\bf 1}_L, \\
K_i= {\bf 1}_1\otimes\dots\otimes {\bf 1}_{i-1}\otimes (s_i^--v_i)\otimes n_{i+1}\otimes{\bf 1}_{i+2}\otimes\dots\otimes {\bf 1}_L,
\eeqn
where ${\bf 1}_i$ denotes the identity operator on site $i$. 
Representing the states $|0\rangle$ and $|1\rangle$ by the column vectors  $(1, 0)^T$ and  $(0, 1)^T$, respectively, the local operators appearing in the above expressions are represented by the matrices: 
\be 
v=
\pmatrix{
1 & 0 \cr
0 & 0  \cr}
\qquad 
n=
\pmatrix{
0 & 0 \cr
0 & 1  \cr}
\qquad 
s^-=
\pmatrix{
0 & 0 \cr
1 & 0  \cr}
\qquad 
s^+=
\pmatrix{
0 & 1 \cr
0 & 0  \cr}
\ee
Let us consider a dual process having the Hamiltonian $\tilde H$ that is related to the $H$ as 
\be 
\tilde H^T=DHD^{-1}.
\ee
It is known that if the operator $D$ is of the form 
$D=\otimes_{i=1}^Ld_i$ with local operators $d_i$ 
represented by the matrix
\be 
d=\pmatrix{
1 & 1 \cr
1 & 0  \cr}
\ee
then, in the case of a symmetric contact process, $\tilde H=H$, 
i.e. the model is self-dual \cite{hv}.
In case of an asymmetric contact process, we obtain by straightforward 
calculations that the dual operators are 
\be
\tilde M_i=M_i  \qquad \tilde  L_i=K_i  \qquad \tilde K_i=L_i, 
\ee
and thus $\tilde H$ describes the asymmetric contact process that differs from the original one in that $\lambda_i$ and $\kappa_{i+1}$ are interchanged.  
The local density $\rho_n(t)$ in case of a fully occupied initial state 
can then be related to the survival probability $P_n(t)$ in the dual process 
in the same way as it has been done 
for the symmetric process in Ref. \cite{hv}. 
For the sake of self-containedness, we recapitulate the calculations here.

The local density can be written in the vector notation as 
\be
\rho_i(t)=\langle s|n_ie^{-Ht}|L\rangle,
\label{rhoH}
\ee 
where $\langle s|$ is the sum of basis vectors, $|s\rangle=\sum_{\eta}|\eta\rangle$, and  $|L\rangle$ denotes the fully occupied state. 
A useful property of $D$ is that it relates the fully occupied state to the empty one: 
\be
\langle s|D^{-1}=\langle 0|,  \qquad D|L\rangle = |0\rangle.
\label{sL}
\ee
Inserting the identity operator $D^{-1}D$ in Eq. (\ref{rhoH}) in the way 
$\rho_i(t)=\langle s|D^{-1}Dn_iD^{-1}De^{-Ht}D^{-1}D|L\rangle$, then using 
Eq. (\ref{sL}) and the relation 
$d_in_id_i^{-1}=v_i-s_i^+$ one arrives at
\beqn 
\rho_i(t)=\langle 0|(v_i-s_i^+)e^{-\tilde H^Tt}|0\rangle \\
= \langle 0|e^{-\tilde Ht}(v_i-s_i^-)|0\rangle \\
= 1 - \langle 0|e^{-\tilde Ht}|i\rangle, 
\eeqn
where $|i\rangle$ denotes the state in which all but site $i$ are empty. 
The 2nd term on the r.h.s. is the probability that, started from a single individual at site $i$ in the dual environment the empty state is reached at time $t$, which is by definition $1-P_i(t)$, thus we obtain Eq. (\ref{rhoP}).

%%%%%%%%%%%%%%%%%%%%%%%%%%%%%%%%%%%%%%%%%%%%%%%%%%%%%%%%%%%%%%%%%%%%%%
\ack
%%%%%%%%%%%%%%%%%%%%%%%%%%%%%%%%%%%%%%%%%%%%%%%%%%%%%%%%%%%%%%%%%%%%%%
The author thanks F. Igl\'oi, I. A. Kov\'acs and G. \'Odor for 
useful discussions. 
This work was supported by the J\'anos Bolyai Research Scholarship of the
Hungarian Academy of Sciences, by the National Research Fund
under grant no. K75324, and partially supported by the European Union and the
European Social Fund through project FuturICT.hu (grant no.:
TAMOP-4.2.2.C-11/1/KONV-2012-0013).

%%%%%%%%%%%%%%%%%%%%%%%%%%%%%%%%%%%%%%%%%%%%%%%%%%%%%%%%%%%%%%%%%%%%%%%
%%%%%%%%%%%%%%%%%%%%%%%%%%%%%%%%%%%%%%%%%%%%%%%%%%%%%%%%%%%%%%%%%%%%%%%

\end{document}